\documentclass{article}

\usepackage{arxiv}

\usepackage[utf8]{inputenc} 
\usepackage[T1]{fontenc}    
\usepackage{hyperref}       
\usepackage{booktabs}       
\usepackage{nicefrac}     
\usepackage{microtype}      
\usepackage{lipsum}
\usepackage{graphicx}
\usepackage{multirow}
\usepackage{amsmath,amssymb,amsfonts}
\usepackage{amsthm}
\usepackage{mathrsfs}
\usepackage[title]{appendix}
\usepackage{xcolor}
\usepackage{textcomp}
\usepackage{manyfoot}
\usepackage{algorithm}
\usepackage{algorithmicx}
\usepackage{algpseudocode}
\usepackage{listings}
\usepackage{natbib}
\usepackage{hhline}
\usepackage{array}
\usepackage{tcolorbox}
\tcbuselibrary{skins}

\title{Towards Usable Privacy Management for IoT TAPs: Deriving Privacy Clusters and Preference Profiles}

\author{
 Piero Romare \\
  Chalmers University of Technology \\
  University of Gothenburg\\
  Gothenburg, Sweden \\
  \texttt{pieror@chalmers.se} \\
   \And
 Farzaneh Karegar \\
  Karlstad University\\
  Karlstad, Sweden \\
  \texttt{farzaneh.karegar@kau.se} \\
  \And
 Simone Fischer-Hübner \\
  Karlstad University \\
  Chalmers University of Technology \\ 
  University of Gothenburg\\
  Karlstad, Sweden \\
  \texttt{simonefi@chalmers.se} \\
}

\begin{document}
\maketitle
\begin{abstract}
    IoT Trigger-Action Platforms (TAPs) typically offer coarse-grained permission controls. Even when fine-grained controls are available, users are likely overwhelmed by the complexity of setting privacy preferences. This paper contributes to usable privacy management for TAPs by deriving privacy clusters and profiles for different types of users that can be semi-automatically assigned or suggested to them. We developed and validated a questionnaire, based on users' privacy concerns regarding confidentiality and control and their requirements towards transparency in TAPs. In an online study (N=301), where participants were informed about potential privacy risks, we clustered users by their privacy concerns and requirements into Basic, Medium and High Privacy clusters. These clusters were then characterized by the users' data sharing preferences, based on a factorial vignette approach, considering the data categories, the data recipient types, and the purpose of data sharing. 
    Our findings show three distinct privacy profiles, providing a foundation for more usable privacy controls in TAPs.
\end{abstract}


\section{Introduction}\label{sec:intro}
The fast proliferation of connected devices and services in the Internet of Things (IoT) has fundamentally transformed how we interact with and experience our environment. IoT-enabled systems offer a wide range of automation benefits, including improved energy efficiency, enhanced safety, and increased convenience for users~\cite{khodabakhsh_2020_home}. However, these systems' ubiquitous data collection and sharing capabilities have also raised significant privacy concerns among users~\cite{zheng_2018_perceptions, naeini_2017_iot}. How users can exercise their right to informational privacy, including control over the collection and use of personal data, should match their expectations~\cite{rao_2020_privacy}. 

This study focuses on Trigger-Action Platforms (TAPs), which enable users to create automated rules that control the behavior of connected devices~\cite{ur_2016_trigger}, allowing for customisation of their everyday digital systems including smart homes, smart cities, and wearables. TAPs host applications that are based on the "if-this-then-that" formula, giving the end-users the role of high-level programmers by connecting different devices and/or services. These platforms, such as IFTTT, Zapier, Power Automate, Make and many others, allow users, even without extensive coding expertise, to create and submit their own applications from user-friendly web interfaces. IoT Trigger-Action applications serve as bridges ("webhooks") connecting smart devices and online services, improving their interoperability. Each app consists of an event-driven program with a code snippet that includes at least a trigger and an action. For example, a user might create a rule that turns on the smart lights when the front door is unlocked.

TAP applications, such as in IFTTT, can collect users' data from a device or service and transmit the data to third party entities involved in the application whenever certain events trigger actions to be performed by these third-party entities~\cite{ifttt_terms_2025}. Significant privacy issues may emerge when users employ this automation and customisation capability with personal or sensitive data collected from IoT devices or inferred by the TAP~\cite{aghvamipanah2024activity}, and then forwarded to other third party entities~\cite{wang2025privacyguard}. Privacy issues also specifically arise, as "overprivilege" is a significant shortcoming of permission models in Trigger-Action IoT Platforms~\cite{fernandes2016security,xu2019privacy,ahmadpanah2023lazytap}.
The permission systems used by many TAPs, including IFTTT, to control these personal data flows are only based on coarse-grained permissions that in most cases provide more generous access permissions and controls than needed and desired by users~\cite{balliu2019securing}.
Nevertheless, even when fine-grained controls are in place, users, for whom privacy and security are usually only secondary goals~\cite{fischer_2024_secondary}, are likely to be overwhelmed with the complexity of setting fine-grained privacy controls that will correspond with their preferences. An abundance of options actually requires more user effort to choose and can leave users feeling unsatisfied with their decisions~\cite{schwartz_2015_paradox}. As concluded in~\cite{balliu2019securing}, fine-grained permission and control systems are needed for IoT TAPs, which provide users with control over how their personal data are used by third-party sites, and additionally these permission and control systems should also be usable. To this end, approaches based on bundling access permissions and controls, and automatically deriving suitable profiles (or ``bundles'') of permissions and controls that align with users' preferences may be beneficial~\cite{balliu2019securing}. Aligning such permission and control systems with users' requirements and data sharing preferences is key to ensuring that these systems are transparent, trustworthy and support the users' expectations regarding control over their personal data. In other words, we need user-tailored privacy permission and control systems~\cite{knijnenburg_2017_UTP}. 

Privacy preferences refer to users' wishes configurations that control how their personal data is shared, accessed, and utilised by others, reflecting their desired level of control over how their data is handled by various entities which is influenced by a combination of factors, including cognitive biases, contextual information, and the trade-offs between privacy and the benefits of sharing data~\cite{acquisti_2015_privacy}. In the IoT context, the privacy preferences are highly variable and context-dependent, as participants may feel more or less comfortable sharing their data based on specific situations or applications~\cite{naeini_2017_iot}. To create a user-tailored privacy permission system, we need simplified privacy management based on personalized settings that align with users' privacy preferences, supporting decision-making more intuitive and less overwhelming. This approach can involve clustering users based on either their privacy behaviors—observable actions they take to protect their data~\cite{lin_2014_preferences,liu_2014_clusters}—or their privacy attitudes, which reflect their internal thoughts and mental states~\cite{kumaraguru_2005_westin,dupree_2016_personas}.

Users often adjust their privacy preferences based on the perceived trade-offs between risks and rewards in specific situations, reflecting an ongoing negotiation between privacy concerns, individuals' perceptions about the implications of sharing information online~\cite{dinev_2006_concerns}--and the benefits of data sharing, as highlighted by~\cite{acquisti_2015_privacy} in their discussion of privacy decision-making dynamics and privacy calculus. However, research shows that users may have misconceptions about security and privacy risks in the context of IoT TAPs. For example, in a previous user study, real IFTTT users expressed that while secrecy and integrity of potential harms were important to them, they believed their applets were safe and did not modify their level of caution even when presented with explanations of potential violations~\cite{cobb_ifttt_2020}. 
Similarly, an online survey study found that users often struggle to understand the security and privacy implications of IoT devices, with many indicating misconceptions about the risks associated with real IFTTT applets, but with guided risk assessments considering the time of day, location, and the presence of others, users were better able to identify potential risks including leakage of sensitive data and unauthorized or unintended access~\cite{saeidi_2022_risk}. Users may also have difficulties with understanding potential privacy risks and how they can set adequate permissions that align with their preferences~\cite{madejski_2012_facebook}, often due to a lack of risk awareness and knowledge about how their data might be used, which further complicates their ability to accurately assess risks~\cite{emami2020informing}.  

Therefore, solely relying on the collected privacy preferences of users who were not well informed about potential risks may not be sufficient for deriving effective controls for protecting users according to their expectations. Since privacy preferences are often shaped by users' perceptions of risk and their privacy concerns~\cite{lee_2017_preference}, the variability in users' risk perception and technological understanding complicates effective decision-making in IoT TAPs. Without sufficient awareness of potential risks, privacy concerns and in turn preferences may not adequately align with the controls needed to safeguard users’ data~\cite{lee_2017_preference}.

Consequently, in this paper, our main goal is to derive clusters of attitudinal privacy concerns and requirements from users who have been exposed to different IoT TAPs scenarios and informed about potential privacy risks. These clusters will then be characterized by users' data sharing preferences to form privacy preference profiles. These profiles can serve as an important step towards deriving suitable bundles of permission settings that better align with users' privacy expectations and needs.

To derive clusters of privacy concerns and requirements, we conducted an online survey using a custom-designed questionnaire. Participants responded to items reflecting three a priori theorized dimensions of IoT-TAP privacy concerns and requirements in the context of four real IoT TAP application scenarios: confidentiality, transparency, and control~\footnote{Measurement validation later indicated that confidentiality and control loaded on a single factor, so subsequent analyses used two dimensions: confidentiality/control, and transparency.}. For each scenario, participants were provided with information about the potential privacy risks associated with that specific application.
Moreover, we used a factorial vignette setting to further describe and characterize these privacy clusters in terms of users' data sharing preferences, focusing on key privacy factors in the context of IoT, as identified by~\cite{iotsecurityprivacy}, to form privacy profiles for IoT TAPs. These key factors include:
(1) data category (personal data) to be shared: personally identifiable information (PII) (name, surname, address, IP address), location data, message and email data, image and video data; (2) purpose of data sharing: main app functionality, personalised app functionality, targeted advertisement;  and (3) data recipient type: with service providers (e.g., parties that are involved in the IoT app as trigger or action providers), government and legal authorities, and (other) third parties. Further details on the factors selection are in Section~\ref{sec:preferences}.

\subsubsection*{Research Questions and Contributions} 
Our survey study was designed to answer the following research questions aimed at achieving our overall research goal for current and potential IoT TAP users with different backgrounds:
\newline \indent \textbf{RQ1}: \textit{What} types of privacy clusters can be identified, and \textit{how}, to serve as a foundation for a usable privacy permission and control management system that reflects concerns and requirements of users who are informed about potential privacy risks in IoT TAP applications?
\newline \indent \textbf{RQ2}: How can these privacy clusters be further characterized based on users' data sharing preferences in IoT TAP applications to form privacy preference profiles? \newline

\noindent Our study contributes to the field with the following key findings that previously have not been investigated in the context of IoT TAPs: 
\begin{itemize}
    \item \textbf{Validated Questionnaire:} We developed and validated a novel questionnaire as a reliable data-gathering procedure to cluster participants, who were informed about potential privacy risks, for four selected IoT TAP application scenarios. 
    \item \textbf{Privacy Clusters for IoT TAP Applications and Users:} From the questionnaire responses, we derived privacy concerns and requirements clusters for participants who were informed about privacy risks in IoT TAP scenarios. 
    \item \textbf{Privacy Preference Profiles:} 
    We showed how the features from our data sharing factorial vignette study procedure matched and characterized the privacy clusters derived and formed privacy preference profiles.
    \item \textbf{Directions for Usable Privacy Management:} 
    We provided insights into and discussed how our results can serve as the first step towards developing usable privacy management systems for IoT TAPs, based on bundles of privacy settings that can be proposed or semi-automatically assigned to users exposed to potential privacy risks, using a simple, validated 6-question questionnaire.
\end{itemize}
To the best of our knowledge, this is the first study to investigate privacy questionnaires, clusters and profiles within the specific context of IoT TAPs. Our research contributions differ considerably from earlier work that focus exclusively on IoT in general, rather than IoT TAPs, because (a) IoT TAPs introduce additional privacy risks through overprivileged permission system, automated actions, and data transfers to third parties, and (b) the privacy factors influencing users’ concerns and preferences for IoT TAP use extend beyond those typically identified for IoT more broadly~\cite{romare_literature_2023, romare_tapping_2023}.

\subsubsection*{Organization}
In Section~\ref{sec:related_work}, we present background about IoT TAP, privacy concerns and preferences and related work on privacy scales and privacy clustering techniques used in other contexts than IoT TAP. In Section~\ref{sec:methods}, we provide the design of our validated questionnaire with the scenarios and risks proposed, address research ethics, and present the recruitment, sample and the statistical methodology that we employed. In Section~\ref{sec:results}, the details about the validity and reliability of the questionnaire are presented, and our research questions are answered. In Section~\ref{sec:discussion}, we discuss the relevance of our results as a step towards implementing usable privacy permission management for IoT TAPs. We also discuss our results in comparison with related work and discuss limitations as well as future research directions. Finally, we conclude this paper in Section~\ref{sec:conclusion} by summarizing the main results of our study and briefly discussing future research directions.
\section{Related Work}\label{sec:related_work}
In this section, we briefly summarise and discuss work related to our research contribution. This includes literature on how to capture privacy concerns and requirements through privacy scales, with a later focus on the context of IoT TAPs, as well as related work on categorising users and deriving privacy clusters and profiles for capturing users' privacy and preferences.

\subsection{Capturing Users' Privacy Concerns and Requirements in IoT}
Privacy concerns, as defined by \cite{smith_1996_information}, refer to the worries and uneasiness of people regarding the loss of privacy and how well they are protected against unauthorised access and misuse of personal information. People were found to be more concerned about their privacy when the information process is not transparent~\cite{phelps_2000_concerns}. 

To explore privacy concerns across various contexts, privacy measurement scales and models have been developed to capture the often latent variable that expresses an individual's privacy attitudes. The Concern for Information Privacy (CFIP) model with 15 items explores organizational privacy practices as well as people's concerns towards four dimensions: collection of data, unauthorised secondary use of data, improper access to data, and errors in data~\cite{smith_1996_information}. As the first validated instrument regarding privacy concerns, the CFIP served as a foundation for the Internet Users' Information Privacy Concerns (IUIPC)~\cite{malhotra_2004_internet}. The IUIPC refines the CFIP, and it is a scale for analysing privacy concerns for internet users that consists of 10 items, or in its shorter version of 8 items~\cite{gross_2021_validity}, among three dimensions such as data collection, control and awareness. The Internet General Privacy Concern (IGPC) scale was developed to understand the e-commerce customers' caution and mistrust over their personal data under the dimensions of information transfer and use of data by~\cite{castaneda_2007_concerns}.~\cite{earp_2005_policies} developed another instrument to verify the alignment between the privacy policies' practices and the users' expectations regarding the employed data protections in online websites considering the personalisation, awareness, data transfer, collection, storage and data access. Regarding online social networks, the multidimensional privacy orientation scale includes four dimensions named privacy as a right, concern about own informational privacy, and concerns about others' privacy. Based on these dimensions,~\cite{baruh2014more} identified three user segments: privacy advocates, individualists, and indifferents. The Value of Other People's Privacy (VOPP) handled how much people protect others' personal information with a validated psychometric scale~\cite{hasan2023psychometric}. The Privacy Attitudes Questionnaire (PAQ) is another instrument to investigate a multidimensional construct considering exposure, willingness to be monitored, interest in privacy and privacy control or trust with 36 items~\cite{chignell2003privacy}.
The Mobile Users' Information Privacy Concerns (MUIPC) was developed by~\cite{xu_2012_muipc} for measuring privacy concerns in mobile applications, including perceived surveillance, intrusion and secondary use of personal information. The MUIPC was later also extended to the IoT context~\cite{foltz_2020_muipciot}. This scale was applied to the IoT to assess comfort levels across health, safety, user experience, and personalisation benefits. In the contexts of user experience and personalisation, comfort levels decreased when data sharing was allowed~\cite{chawdhry_2022_concerns}, while providing transparent information can often ease concerns~\cite{magrizos2025transparency}.
In the context of IoT TAPs,~\cite{romare_tapping_2023} used focus groups as a qualitative research method for eliciting privacy factors that impact users’ concerns and preferences for using IoT TAPs, and identified transparency, control, risks, trust and confidentiality as relevant factors. 
~\cite{saeidi_2022_risk} were the first to use a quantitative approach for measuring users’ concern scores for using IFTTT applets and explored if and how those concerns were impacted by different contextual factors such as location, time of the day and other people's access to the trigger or action service. Their results showed that concern scores were low when their participants were only exposed to the app descriptions, which indicated a limited awareness of potential risks. Considering the smart home environment, in~\cite{apthorpe_2018_ci}, the acceptability of thousands of information flows that align with indications of IFTTT user studies has been measured. Indeed, individuals' priorities are user consent and transparency in data collection practices since user awareness and control over data sharing are key factors influencing privacy concerns. 
While influential in the design of our questionnaire in terms of selection of items and dimensions we investigated, none of these earlier studies, in contrast to our work, focused on capturing privacy clusters and profiles in the IoT TAP context. Further details about the design, selection and modification of existing and novel items included in our questionnaire are discussed in Section~\ref{sec:concerns}.

A questionnaire can be used as a measurement tool that can collect data to perform the clustering in this regard when properly evaluated~\cite{biselli_questionnaire_2022}. Thus, we validated a privacy concerns and requirements questionnaire developed by adapting existing items to the context of IoT TAPs and proposing a tool to measure privacy attitudes specific to these technologies and show how this questionnaire can be used to cluster study participants who have been informed about potential privacy risks.

\subsection{Privacy Categorization of Users, Privacy Clusters and Profiles}
Previously,~\cite{inverardi_2023_categorisation} presented a systematic literature review on privacy categorisation, including work related to our study, and showing how terminologies and methods evolved over time, concluding with discussing the potential of these approaches to support users in managing their privacy with the help of recommendation systems that can be developed. The term segmentation can be essentially expressed as \textit{a model-driven procedure of partitioning a dataset or extracting associated features}. The term clustering is usually used to refer to the \textit{data-driven mathematical process of grouping similar data points}. The term profiling can generally be specified as a \textit{hybrid approach that combines multiple data sources and can involve statistical analysis to identify characteristics of individuals or groups}. Personas, on the other hand, refer to \textit{attributing new parameters to existing segments or clusters}.

Early work on privacy categorisations of users dates back to 1990/1991 when Alan Westin proposed his first privacy segmentation from responses to four questions. His approach grouped individuals into three categories: privacy fundamentalists, pragmatists, and the unconcerned~\cite{westin_2003_bibliography}. His methodology provided a framework for segmenting participants based on their privacy concerns and attitudes and his segmentation has been utilised by academics in a wide range of domains for analysing and categorising users regarding their privacy attitudes and concerns, but was also discussed for its limitations~\cite{kumaraguru_2005_westin}. Individuals' privacy strategies from Westin's key aspects of privacy, applied in both analogue and digital contexts, are the right to be alone, intimacy, anonymity, and limited information disclosures~\cite{westin_1968_privacy}. These were refined to address more specific IoT privacy necessities by~\cite{ziegeldorf_2014_privacy}, including awareness of privacy risks posed by connected devices and services, control over personal data collection and processing, and awareness and control of third-party data sharing.

~\cite{inverardi_2023_categorisation} highlight several key changes as a result of their review on privacy categorization and one significant shift mentioned is the evolution from a focus on data minimization and binary privacy choices to a more nuanced understanding of privacy preferences modelled using qualitative, quantitative and hybrid approaches, where mostly three categories of users were captured. Early research often presented privacy decisions as static and simplistic, but the rise of complex digital ecosystems, such as IoT and social media, required a more dynamic approach, where users’ privacy preferences vary based on the specific context or scenario. Building on this evolution, our study similarly acknowledges the importance of context dependency in privacy preferences. Accordingly, in our questionnaire, we incorporated scenario-specific considerations to capture the variability in users' privacy concerns and requirements in different IoT TAPs scenarios, as described in Section~\ref{sec:expertevaluation}.

Clustering algorithms, such as hierarchical clustering and K-means have been used to derive privacy profiles. In~\cite{brandao_2022_clustering}, three privacy profiles were extracted from the hierarchical clustering algorithm, whereas six more granular and nuanced profiles emerged from the k-means approach for Android smartphones, considering detailed app category–permission combinations in a vector meticulously representing each individual user. 
Privacy preferences can also be modelled and used as input data to train machine learning models for clustering and predicting users' privacy settings in the form of profiles, thereby helping to address the cold start problem. For example,~\cite{bahirat_2018_preferences} tested attitude-based, fit-based (achieving 82\% accuracy with three profiles), and agglomerative clustering solutions. 
Their analysis demonstrates how participants were more comfortable when information where not continuously shared, and instead perceived more risk when the recipient was unknown using the dataset provided by~\cite{lee_iot_2016}. Originally, that dataset was analyzed for clustering IoT users by exploring the relationships between where data are collected, what type of data are gathered, who receives them, for what purposes, and how frequently collections occur in IoT scenarios. That analysis resulted in four distinct clusters of users based on their privacy preferences related to notifications, permissions, comfort, risk, and appropriateness~\cite{lee_iot_2016}. 
Another extensive study in IoT, involving over a thousand participants, used factors such as who, what, purpose, storage location, and action to create data sharing scenarios for users of household IoT devices. With the help of machine learning algorithms, "smart" profiles were derived considering the adoption decision and the contextual factors~\cite{he_iot_2019}. 
In the smartphone context, privacy profiles were used to implement a privacy assistant, which was, for instance, developed by~\cite{liu_recommendation_2016} to help Android users manage their privacy preferences with a behavioral field study. The privacy assistant could recommend privacy settings thanks to a Support Vector Machine classifier that reached 79\% of acceptance rate.
In another study, four privacy profiles were created considering the number of applications installed and users' permissions decisions from their phones and these four clusters were then combined with self-reported privacy attitudes and intentions of use to derive profiles~\cite{alsoubai_2022_smartphone}.

To the best of our knowledge, we are the first to integrate, in the context of IoT TAPs, attitudinal privacy concerns and requirements clusters collected with a questionnaire and modelled into profiles combining the data sharing preferences. By bridging self-reported privacy concerns with context-specific choices, our work can lay the foundation for future studies using behavioral data and aims to contribute to the development of usable privacy management and privacy assistants for IoT TAPs. As we will further elaborate in Section~\ref{sec:disc-noverlty}, our derived clusters not only differ domain and scope-wise but also content-wise from other clusters derived in earlier related work.
\section{Method}\label{sec:methods}
\begin{figure}[ht]
    \centering
    \includegraphics[width=\linewidth]{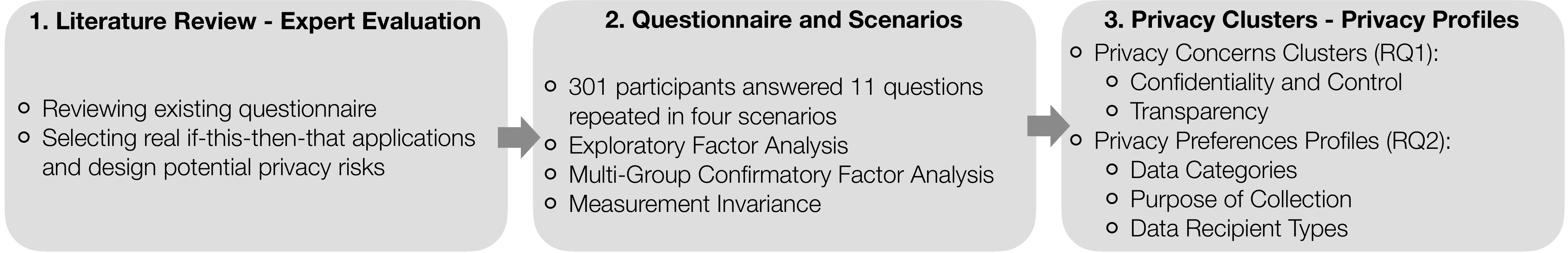}
    \caption{The Main Steps used in this study}
    \label{fig:study_design}
\end{figure}

In this section, we present our approach to address the research questions presented in Section~\ref{sec:intro}. First, we conducted a literature review and an expert evaluation to set the scene and define the TAP scenarios for our questionnaire. Then, we conducted our questionnaire to capture users' attitudinal privacy concerns and requirements across three key dimensions: transparency, control, and confidentiality. To ensure the effectiveness of our questionnaire in extracting meaningful privacy clusters, we demonstrate the validity and reliability of this measurement tool. Lastly, with a factorial vignette study approach, we characterized the derived clusters based on users' data sharing preferences, resulting in distinct privacy profiles. The steps we followed are depicted in~\autoref{fig:study_design}. In addition, we discuss the ethical considerations, the participant recruitment process, and our data analysis methods in this section. 

\subsection{Expert evaluation and IoT TAP Scenarios}\label{sec:expertevaluation}
To capture users' attitudinal concerns and requirements related to IoT TAPs, we need a clearly defined context that frames the scenario in which users can respond to privacy-related questions. Consequently, we defined four scenarios that each presents the application's description and the privacy risk related to additional information that an adversary could infer about the user.

Information about privacy risks was included, since, as introduced in Section~\ref{sec:intro}, a gap exists in participants' awareness of the risks associated with additional information that could be inferred~\cite{alqhatani_2023_inference} or, in other words, if there's a leak of data, which could also be due to a possible IFTTT application's violation as exemplified in~\cite{surbatovich_2017_risks}. The purpose of introducing the scenarios is to give examples of existing IoT TAP applications with related risks and help the participants understand how the applications can be used in their everyday lives.

\begin{table}[ht]
    \caption{Scenario Figure Reference (\#) with description and privacy risk.}
    \centering
    \renewcommand{\arraystretch}{1.1} 
        \begin{tabular}{|>{\centering\arraybackslash}p{0.5cm} 
                        |>{\arraybackslash}p{4cm} 
                        |>{\arraybackslash}p{7cm}|}
        \hline
        \textbf{\#} & \textbf{App Description} & \textbf{Privacy Risk} \\ \hline
        
        \ref{fig:scenario1} & 
        If a new sleep is logged by your smartwatch, then add an event in your cloud calendar with sleep information. & 
        If an adversary has intercepted the communication, there is a risk that sleep routine details may be leaked and used to infer a person's stress levels, fatigue, or suggest sleeping pills. \\ \hline
        
        \ref{fig:scenario2} & 
        If your location is outside your home, then lock the door. & 
        If an adversary collects your location information, they could predict when you won’t be at home. \\ \hline
        
        \ref{fig:scenario3} & 
        If a new photo is in your smartphone's camera roll, then upload it to your cloud storage. & 
        If an adversary gains unauthorized access to the cloud storage, they could potentially extract sensitive information from the uploaded photos. This may include a collection of places you visited, their time, and people with you, leading to a privacy breach and the unintentional exposure of personal details. \\ \hline
        
        \ref{fig:scenario4} & 
        If you like a video, then post it in your online social network account. & 
        Posting content to an online social network may allow it to target you with advertisements based on your personal preferences from the media platform. \\ \hline
        
        \end{tabular}
    \label{tab:scenarios}
\end{table}
 
As a first step for selecting suitable applications for the scenarios, we compared the dataset available in~\cite{kalantari_ifttt_2022} and the most used  IFTTT applications based on their categories~\footnote{https://ifttt.com/content\_map (visited Nov 2023)}: 1) mobile, devices and accessories, 2) news and information, 3) social media, 4) notifications 5) business tool 6) photo and video.
We manually filtered and selected applications that included personal data flows that allowed us to infer sensitive information, and then categorized them into four categories: mobile devices and accessories, social media, business tools, and photos and videos. This first step resulted in 16 realistic scenarios that use common IoT devices or services. 

We then conducted semi-structured interviews with six experts to form and filter out our scenarios and to gather expert opinions on potential privacy risks and their severity for each scenario. The results of our expert interviews finally led to the selection of four scenarios to be included in our study. Our expert participants were PhD students and post-doctoral researchers with backgrounds in Information Security and Privacy at Chalmers University of Technology who volunteered to take part in the interviews after our call for expertise requirements and provided their informed consent before the interview started. The interviews took 25 minutes on average and included the following steps: 1) introducing the context of information flow in IoT TAPs, 2) explaining the goal of the current study, and 3) exposing participants to 16 scenarios and asking them to identify potential risks.

\begin{figure}[ht]
  \centering
  \includegraphics[width=\linewidth]{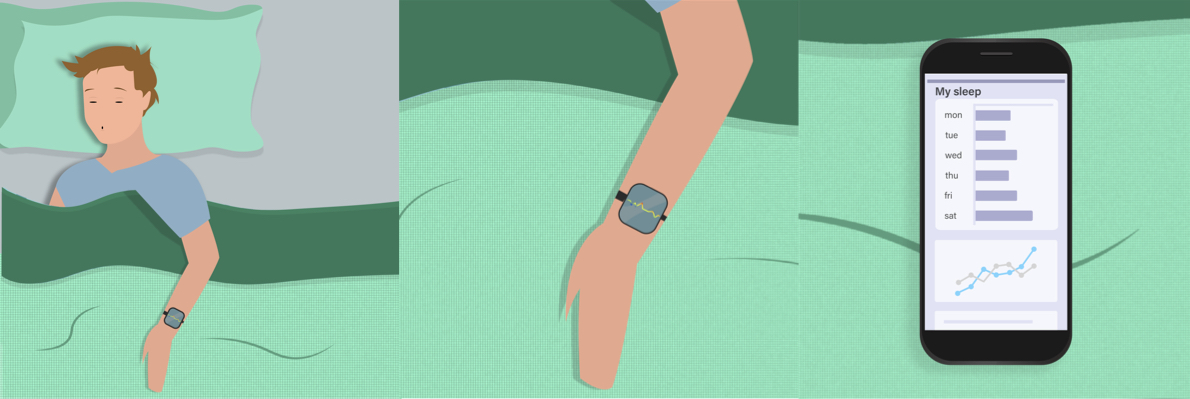}
  \caption{Scenario \#1: if new sleep log, then upload sleep information on cloud calendar - ID: wsTcJyNt}
  \label{fig:scenario1}
\end{figure}

\begin{figure}[ht]
  \centering
  \includegraphics[width=\linewidth]{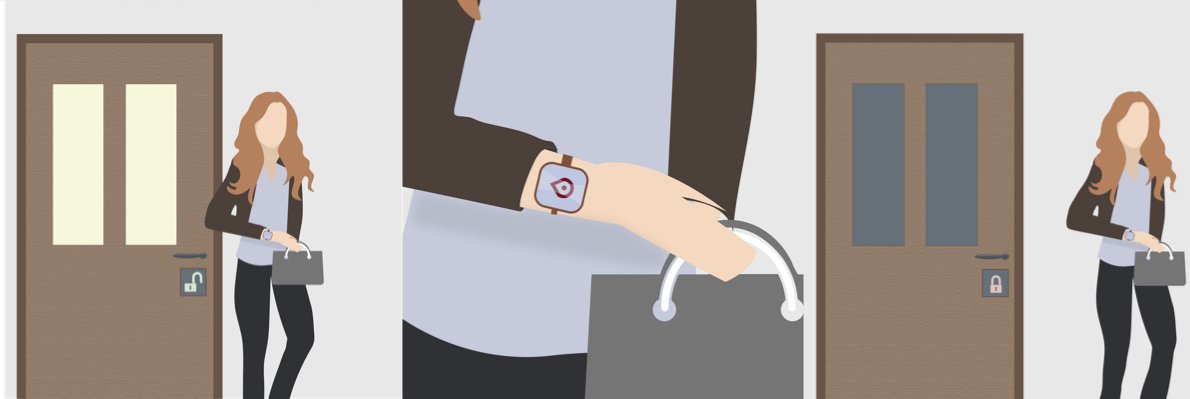}
  \caption{Scenario \#2: if outside home, then lock the door - ID: NBwbDaze}
  \label{fig:scenario2}
\end{figure}

\begin{figure}[ht]
  \centering
  \includegraphics[width=\linewidth]{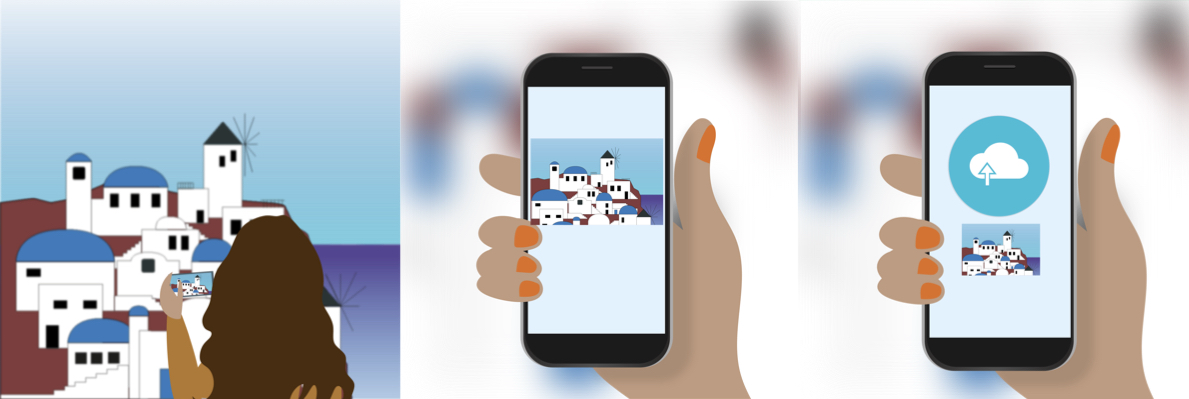}
  \caption{Scenario \#3: if new photo, then upload on cloud storage - ID: th9yp6nk}
  \label{fig:scenario3}
\end{figure}

\begin{figure}[ht]
  \centering
  \includegraphics[width=\linewidth]{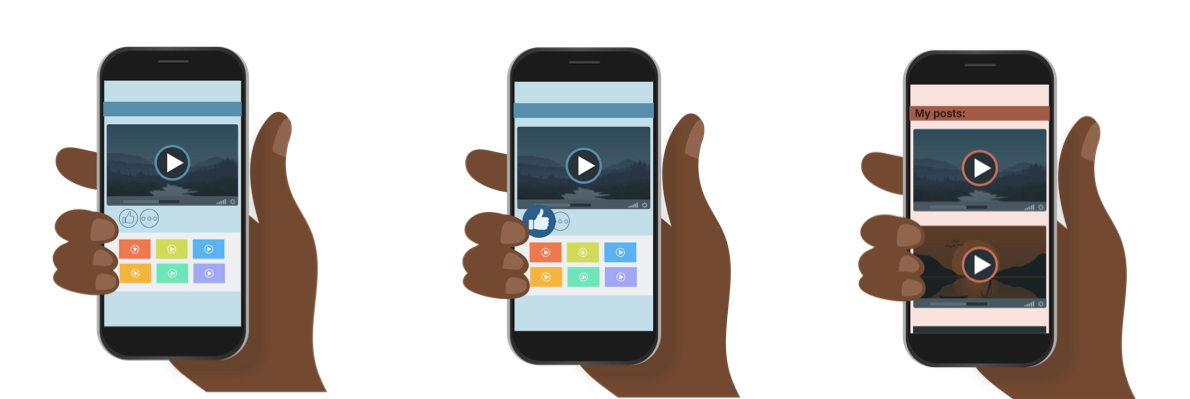}
  \caption{Scenario \#4: if you like a new video, then post it on online social media - ID: k6LpJhGg}
  \label{fig:scenario4}
\end{figure}

The privacy risks for each scenario can have personal implications through privacy violations that may cause embarrassment or leak behavioral data~\cite{surbatovich_2017_risks}. The results of these interviews led to the selection of a suitable subset of if-this-then-that IoT TAP scenarios with certain characteristics, such as personal data flow, to be used in our study (see~\autoref{tab:scenarios}). The scenarios are related to real IFTTT applications that can be 
accessed at their website~\footnote{https://ifttt.com/applets/(id)} using the id provided under each Figure representing the application. To help our participants better grasp the scenarios and their context, each scenario was accompanied by a description and related figure (see~\autoref{fig:scenario1},~\autoref{fig:scenario2},~\autoref{fig:scenario3},~\autoref{fig:scenario4}).


While potential privacy risks that were identified via our expert interviews may be obvious for privacy and security experts, earlier research by leading privacy scholars, including~\cite{acquisti_2015_privacy} and~\cite{Solove2013}, shows that lay users often misjudge and underestimate privacy risks and potential harms, and fail to connect abstract risks to concrete consequences. Also, as pointed out above,~\cite{saeidi_2022_risk} demonstrate that many people are not aware or have misconceptions about privacy risks in the context of IoT TAPs. This motivated us to inform test participants about potential privacy risks and consequences related to the IoT TAP that were identified by our experts.

\subsection{Questionnaire Design}\label{sec:design-survey}
In the recruitment invitation message, we briefly introduced the context of the questionnaire to the participants. We provided the Wikipedia definition of IoT and introduced IoT Trigger Action platforms with an example: "If enter a room, then turn on the lights". In addition to a self-explained image which described the data flow between the two entities, we added text details about the trigger, the action and the interconnection between them. If the participant decided to continue and take part in the study, the consent form and the information were provided to the participant. Upon giving their consent, they answered a few demographic questions and questions related to their usage of IoT devices. We asked about the region of residence (EEA / US), the age group (18-25 / 26-35 / 36-45 / 46-55 / 55+), the gender (Woman, Man, Non-binary, Prefer not to say), how many IoT devices they owned (0 / 1-2 / 3-4 / 5 or more), if they had data breach experiences (Yes / No), whether they used or heard of any IoT TAPs (Yes / No) and lastly their highest level of education (Less than High School / High School / University Degree). After these questions, participants were exposed again to the introduction to IoT, IoT TAP, and IFTTT with an example. The core part of our questionnaire then followed, which included two main parts for deriving privacy concerns and requirements, and data sharing preferences.

In the first part, we presented four scenarios in the context of IoT TAP to each participant and after each scenario, they answered 11 question using a 5-point Likert scale (from strongly disagree to strongly agree) for capturing attitudinal privacy concerns and requirements related to three dimensions of confidentiality, control, and transparency (presented in~\autoref{tab:questionnaire} and in~\autoref{app:11_questions}). In the second part, we captured users' data sharing preferences in the context of IoT TAPs. Participants had to indicate their willingness (yes or no) to adopt/accept 36 specific IoT TAP scenarios which varied in three aspects: 1) the data category to be shared, the purpose of data sharing, and the data recipient type (see~\autoref{tab:practices} and~\autoref{app:factorial_vignette}). 

\subsubsection{Privacy Concerns and Requirements}\label{sec:concerns}
We designed a questionnaire that captures attitudinal privacy concerns and requirements in three dimensions related to confidentiality of personal data, user control, and transparency for our defined IoT TAP scenarios. A systematic literature review analyzed the developments of 16 privacy concern scales and concluded that there is a lack in deriving the constructs from existing theoretical frameworks to get common definitions~\cite{bartol_2021_scales}. Our three dimensions correspond to three dimensions of the six-axis privacy protection goals framework's taxonomy presented by~\cite{hansen_privacy_2015}. These three data protection goals play an important role in the design of a transparent permission control system, which allows users to protect and control the disclosure of and access to their personal data. Anonymity, secrecy and confidentiality were shown as significant contributors to the levels of control over users' personal data~\cite{dinev_privacy_2013}. When users can control and manage their data, they have fewer privacy concerns~\cite{xu_2011_control,dinev_privacy_2013}. Transparency is a key aspect for building trust between the users and those who are processing their data~\cite{xu_2024_transparency}. Other important data protection goals of unlinkability/anonymity or data minimisation of the taxonomy by~\cite{hansen_privacy_2015} can be more effectively protected by a privacy-preserving TAP re-design as suggested in~\cite{ahmadpanah2023lazytap} rather than by permission controls, and were therefore not directly captured by our questionnaire.

Basing our questionnaire on these three dimensions was also motivated by earlier work conducted by~\cite{romare_tapping_2023} that identified transparency, control and confidentiality as important privacy factors playing a role in users’ concerns and requirements for using IoT TAPs and its automation function. In their qualitative user study, participants particularly demanded control over the final action, expected to be able to restrict or disallow full automation, and required transparency of automation settings, rules, and data recipients, as well as the confidentiality of communicated and stored data. 

When possible, a facilitating practice to get empirical evidence from a scale would be to reuse, combine or calibrate the items of an existing one~\cite{preibusch_2013_reuse}. Our questions, related to these three privacy dimensions, were designed by examining the literature on similar constructs applied in the general privacy context. Such selections were partially sustained by an existing literature review~\cite{romare_literature_2023} within the context of if-this-then-than applications. The confidentiality dimension was developed by carefully selecting questions from~\cite{cobb_ifttt_2020,dinev_privacy_2013,maus_privacy_2021} that belong to definitions such as integrity, secrecy, confidentiality, and secondary use of personal information. The transparency dimension was entirely developed by Awad and Krishnan to measure the customers' perceived information transparency in online personalized services and advertising~\cite{awad_privacy_2006}. However, their instrument may not be applicable directly to the IoT context. Therefore, we performed substantial updates on the items (see~\autoref{tab:questionnaire-inspired}) as practice done in~\cite{dinev_privacy_2005} considering the IoT Trigger-Action platforms and the presentation format to our participants with the scenario descriptions and images (\autoref{fig:scenario1},~\autoref{fig:scenario2},~\autoref{fig:scenario3},~\autoref{fig:scenario4}).

\begin{table}[ht]
    \caption{Study part 1: Questionnaire related to privacy concerns and requirements repeated per scenario with Dimensions and Items.} 
    \centering
    \renewcommand{\arraystretch}{1.2} 
        \begin{tabular}{|>{\centering\arraybackslash}p{2.25cm} 
                        |>{\centering\arraybackslash}p{0.5cm} 
                        |>{\arraybackslash}p{8.5cm}|}
        \hline
        \textbf{Dimension} & \textbf{id} & \textbf{Item} \\ \hline
        \multirow[c]{4}{*}{\textbf{Confidentiality}} 
            & u1 & I feel that as a result of using this app, my personal information may not remain confidential~\cite{dinev_privacy_2013} \\ \hhline{~--}
            & u2 & I am concerned that this app may sell my information to other companies or institutions without my permission~\cite{maus_privacy_2021} \\ \hhline{~--}
            & u3 & I believe that if I use this app, unauthorized people will have access to my data~\cite{dinev_privacy_2013} \\ \hhline{~--}
            & u4 & I know that by using this app, an incidental data leakage may happen~\cite{cobb_ifttt_2020} \\ \hline
        \multirow[c]{4}{*}{\textbf{Control}} 
            & c1 & I would be upset if this app unintentionally triggered and processed my data~\cite{cobb_ifttt_2020} \\ \hhline{~--}
            & c2 & I am concerned that I may lose control over my data by using this app~\cite{cobb_ifttt_2020} \\ \hhline{~--}
            & c3 & I feel I have control over my data in this app if the data collection happens in compliance with the Privacy Policies, Rules, and Standards~\cite{awad_privacy_2006} \\ \hhline{~--}
            & c4 & For this app, I don't want it automatically running; I prefer to press a button before the action runs~\cite{romare_tapping_2023} \\ \hline
        \multirow[c]{3}{*}{\textbf{Transparency}} 
            & t1 & It’s important for me to know what information the service companies involved in this app store about me in their database~\cite{awad_privacy_2006} \\ \hhline{~--}
            & t2 & For this app, I want to receive a summary about the data processing that occurs (e.g., how my data are manipulated to produce meaningful information)~\cite{awad_privacy_2006} \\ \hhline{~--}
            & t3 & It’s important for me to know if the data from this app will be sold to third-parties for marketing purposes~\cite{awad_privacy_2006} \\ \hline
        \end{tabular}
    \label{tab:questionnaire} 
\end{table}

\begin{itemize}
    \item \textbf{Confidentiality} addresses potential unintended disclosure through automated data transfers between trigger and action services.
    \item \textbf{Control} addresses the preference to initiate actions manually rather than having them run automatically.
    \item \textbf{Transparency} addresses desire about information collected and shared in cross-service data flows and sharing with third parties.
\end{itemize}

In ~\autoref{tab:questionnaire}, we provide the details of each item and the source from which they were initially derived. The items (statements) of the questionnaire were repeated four times, once after each scenario. Both the scenarios and questionnaire items were shown to participants in random order and answered by a 5-point Likert Scale from "strongly disagree" to "strongly agree" (e.g. a participant answers the questions in random order 
for each of the four scenarios, which appeared in a random order for each participant).

\subsubsection{Data Sharing Preferences}\label{sec:preferences}
The second part of our questionnaire is composed of features which describe a context that can have an impact on users' decisions for sharing their data, and thus can be important parameters for a privacy permission management system or access control system. These features serve as independent variables in our factorial vignette study: data category, the purpose of data sharing, and data recipient type. We chose these factors as they represent core elements for understanding the context of personal data processing and thus for judging the sensitivity of personal data, which is highly context-dependent (see also Census Decision~\cite{BVerfG1983CensusVerdict}). In addition, Art. 13 of the EU General Data Protection Regulation (GDPR) and other privacy laws require privacy notices to inform data subjects at least about these factors. These three factors are also directly aligned with the core structure of the IoT Privacy Label framework~\cite{iotsecurityprivacy}, which defines them as essential for informing users about how their data is handled. Our factorial vignette study (4x3x3) was performed using the related sub-levels of each feature as shown in~\autoref{tab:practices}.

\begin{table}[ht]
    \caption{Study part 2: factorial vignette study related to data sharing preferences.}
    \centering
    \renewcommand{\arraystretch}{1.15} 
        \begin{tabular}{|>{\centering\arraybackslash}p{4cm} 
                        |>{\centering\arraybackslash}p{8cm}|}
        \hline
        \textbf{Features} & \textbf{Sub-levels} \\ \hline
        
        \multirow{4}{*}{\textbf{Data Category}} 
            & Location Data \\ 
            & Image and Video Data \\ 
            & Personal Information \\ 
            & Message and Email Data \\ \hline
        
        \multirow{3}{*}{\textbf{Purpose of Data Sharing}} 
            & Main app functionality \\ 
            & Personalised app functionality \\ 
            & Targeted Advertising \\ \hline
        
        \multirow{3}{*}{\textbf{Data Recipient Type}} 
            & Third Parties \\ 
            & Government and Legal Authorities \\ 
            & Service Providers \\ \hline
        
        \end{tabular}
    \label{tab:practices}
\end{table}

Each participant answered a total of 36 yes-no questions, in a random order, in the form of: "Would you accept the following for running an IoT application? Your \{data category\} is shared for \{purpose of data sharing\} and with \{data recipient type\}". 

\subsection{Participants and Ethics}\label{sec:demographic}
The study was conducted in compliance with the EU General Data Protection Regulation (GDPR) and the ethical review act of Sweden. The study design was reviewed and accepted by the data protection officer of Chalmers University of Technology and Karlstad University's ethics advisor. We asked for and obtained informed consent from the six experts we interviewed to design the study and from the participants of our online questionnaire (before they took part in the study). All collected data has been securely pseudonymized and is protected following the rules of the GDPR. We selected LimeSurvey as a survey tool because it has an EU-based (Germany) data controller with data processing policies that comply with the GDPR. 

\begin{table}[ht]
    \caption{Demographics from our participants sample.}
    \centering
    \renewcommand{\arraystretch}{1.2} 
    \begin{tabular}{|
        >{\centering\arraybackslash}m{3cm}|
        >{\centering\arraybackslash}m{3cm}|
        >{\centering\arraybackslash}m{2.5cm}|
        >{\centering\arraybackslash}m{2.5cm}|}
        \hline
        \textbf{Gender} & \textbf{Age} & \multicolumn{2}{c|}{\textbf{Education}} \\ \hline
        Female 48\% & 18-25 32\% & \multicolumn{2}{c|}{Less than High School 1\%} \\
        Male 50\% & 26-35 33\% & \multicolumn{2}{c|}{High School 34\%} \\
        Non-binary 1\% & 36-45 17\% & \multicolumn{2}{c|}{University Degree 65\%} \\
        Prefer not to say <1\% & 46-55 11\% & \multicolumn{2}{c|}{} \\
         & 55+ 7\% & \multicolumn{2}{c|}{} \\ \hline
        \textbf{Region of Residence} & \textbf{\# Devices} & \textbf{TAP} & \textbf{Breach} \\ \hline
        EEA 50\% & 0 18\% & Yes 20\% & Yes 37\% \\
        United States 50\% & 1-2 43\% & No 80\% & No 63\% \\
         & 3-4 25\% & & \\
         & 5+ 14\% & & \\ \hline
    \end{tabular}
    \label{tab:demographics}
\end{table}

We recruited a total number of 301 participants from the Prolific platform for our study (see~\autoref{sec:design-survey}). We selected a balanced distribution between males and females ($\approx$ 50\% - 50\%) in the European Economic Area (EEA) and the United States (US) as western regions, where IoT TAPs are more diffused~\footnote{see also: https://6sense.com/tech/integration/ifttt-market-share (visited Nov 2023)}. 
Our gender-balanced sample was weighted towards younger participants, which we, however, considered as well acceptable, as current and future IoT TAP users should typically be younger. Moreover, we chose to include both participants who are using TAPs already, as well as participants who are not using IoT TAPs yet. Participants who use IoT TAPs already could likely be mostly classified as early adopters, who often tend to be "privacy unconcerned"~\cite{gollust2011motivations}. Therefore, we opted for not involving just the current users of IFTTT since they could introduce a bias in our study towards a less concerned user population, their extroversion that impacts information sharing~\cite{lynn_2017_social} and different perceptions of convenience~\cite{lafontaine_2021_attitude} - and future users might not have been well represented. 

Participants were compensated 11£ / hour for completing the study, with a median time required of 10 minutes. Because of the relatively quick time of completing the questionnaire, we did not include any attention check questions, but instead, we combined the completion times with outlier analysis and Cronbach's $\alpha$~\cite{meade_2012_responses} in answering the 11 questions in each scenario in the first part (see Section~\ref{sec:concerns}) and the total time spent in answering the 36 yes/no questions in the second part related to the factorial vignette study (see Section~\ref{sec:preferences}). No outliers were detected with the Interquartile Range (IQR) rule.  

\subsection{Statistical Testing}
For our questionnaire's reliability and validation, and to derive clusters, we adapted the procedure outlined in~\cite{faklaris_sa6_2019,buchi_cfa_2017,deng_2005_multi}. First, we started with Exploratory Factor Analysis (EFA). After that, we investigated the validity (whether the questionnaire's items actually measure what we propose), reliability (whether the questionnaire's items are consistent), and global fit (how collectively the items are performing) of our questionnaire, by employing the Multi-Group Confirmatory Factor Analysis (MGCFA). Lastly, we implemented hierarchical clustering to obtain groups of participants to characterize the privacy concerns and requirements in clusters and later with data sharing preferences in profiles.

\subsubsection{Questionnaire Validation and Reliability}
The EFA is a data-driven methodology which provides relations between the items in the questionnaire and it is relevant for developing and refining the questionnaire instrument~\cite{conway_2003_efa}. It gives the direction of potentially aggregating more than one item under a dimension. To verify the EFA results which discern the factors accounting for the correlation between observed variables without necessitating underlying theoretical frameworks~\cite{reio_2015_efa,byrne_sem_2005}, it is relevant to perform the Confirmatory Factor Analysis (CFA). 

The CFA is essential to establish reliability and validity to ensure that the constructs assessed by the questionnaire accurately represent the theoretical concepts. When, as in our study, more groups or conditions are employed, the Multi-Group Confirmatory Factor Analysis (MGCFA) is suitable to ensure that the questionnaire is measured equivalently across groups or conditions. It has been used in technology acceptance model studies to measure the invariance across populations~\cite{saritepeci_2024_mgcfa, li_2021_mgcfa} and contexts~\cite{bansal_2016_multi}. By testing the measurement invariance, we can verify that the questions are valid and with the same meanings in the four scenarios we proposed. Such a test implies five steps for demonstrating the full invariance among scenarios: 1) fit CFA models independently one from the others (baseline models); 2) fit MGCFA without constraining the factor loadings estimated (configural model); 3) fit the MGCFA by constraining the factor loadings to be the same among the four scenarios (metric model); 4) fit the MGCFA by constraining the factor loadings and the intercepts to be equal across the four scenarios (scalar model); 5) fit the MGCFA by constraining the factor loadings, the intercepts and the residuals to be equal across the four scenarios (full model). The demonstration of the measurement invariance process was finalized by comparing the Comparative Fit Index (CFI) between the ones obtained in step 2, step 3, step 4 and step 5 which is supported when the difference is $\leq 0.01$~\cite{chen_2010_cfi}. When demonstrated, the measurement invariance ensures that the dimensions hold across different conditions, so measured equally across them~\cite{putnick_2016_measurement}, allowing merging the questionnaire data from each scenario to perform the clustering. The overall performances of the MGCFA models are evaluated with global fit and by verifying the validity and reliability~\cite{gross_cfa_2023} considering Cronbach's $\alpha$ ($\alpha > 0.7$), Root Mean Square Error of Approximation of (RMSEA $< 0.08$), Standardized Root Mean Square Residual (SRMR $< 0.08$), Comparative Fit Index (CFI $> 0.95$), Tucker-Lewis Index (TLI $> 0.95$), Average Variance Extracted (AVE $> 0.5$), the HeteroTrain-MonoTrait (HTMT $< 0.85$) and the Congeneric Reliability ($\omega > 0.7$).

\subsubsection{Extraction of Privacy Clusters and Profiles}
Unsupervised learning is a branch of machine learning that includes clustering and dimensionality reduction algorithms. Hierarchical clustering is an agglomerative clustering algorithm and it is built in a tree form as a dendrogram. In agglomerative clustering, each data point starts as its own cluster, and clusters are merged iteratively considering the distance between those data points using distance metrics. Previous works employed hierarchical clustering to derive privacy clusters and then profiles to manage privacy preferences~\cite{lin_2014_preferences}. Unlike algorithms such as K-means, hierarchical clustering does not require a pre-defined number of clusters. Instead, the number of clusters can be determined by cutting the dendrogram at a specific threshold or by evaluating the silhouette score~\cite{shahapure_2020_silhouette}, ranging from $[-1,+1]$. The dendrogram, created in the hierarchical clustering procedure, shows the levels of similarity between the clusters and allows a visual representation of how the participants, in our context, relate to their concerns and requirements. 

K-means, another popular algorithm~\cite{liu_2014_clusters}, requires the number of clusters to be specified in advance, which can influence the clustering outcome. Other methods used, and often combined, for such a quantitative analysis are principal component analysis, latent semantic analysis and non-negative matrix factorization~\cite{salminen_persona_2020}. Despite the hierarchical clustering being more computationally expensive~\cite{abdalla_2022_complexity}, we chose to employ it due to the relatively small size of our dataset and the interpretability benefits. 

To derive profiles, as defined in Section~\autoref{sec:related_work}, the clusters required multiple data sources to be combined based, for example, on correlated data~\cite{inverardi_2023_categorisation}. To this end, we describe the clusters based on privacy concerns and requirements with the data-preference features collected from the factorial vignette study procedure. The goal of the factorial approach is to gain a better understanding of an individual's judgement and decision-making by using scenarios that consist of features that vary according to their associated (sub)levels and transforming the clusters in profiles. The sub-levels are often presented in the context of a scenario, which serves to situate the survey participant in a specific context. In our work, we are focused on the isolation of those parameters, in terms of allowing or denying potential adoption, regarding their combinations.
 \section{Results}\label{sec:results}
In this section, we present the results of our analysis. The first step is to ensure that our questionnaire is a valid and reliable measurement tool with the EFA and MGCFA. We ran the MGCFA, since our study setup implied the repetition of the same questionnaire in four scenarios that belong to real IFTTT applications. The second step is to derive the privacy concerns and requirements clusters per scenario. Lastly, we used the privacy clusters as ground-truth labels to improve their characterization with the factorial vignette study procedure related to data sharing preferences.

\subsection{EFA and MGCFA of our Questionnaire}
We start our analysis by verifying the feasibility of running the EFA using the Kaiser-Meyer-Olkin factor adequacy (KMO). We ran independently such analysis for all four scenarios which resulted to be with a score of $0.90$ for Scenario \#1, $0.89$ for Scenario \#2, $0.90$ for Scenario \#3, and $0.90$ for Scenario \#4. We then run the EFA using the polychoric correlation matrix and the \textit{oblimin} rotation method. We evaluate the EFA considering the factor loadings that are recommended to load with a threshold of $>0.32$ and they should be dropped if the related cross load with more than one factor is above this threshold~\cite{tabachnick_2013_crossload} and when the commonalities are $h^{2}\geq 0.40$~\cite{costello_2005_efa}. This exploration shows the possible structures of the model. 
The evidences suggested a transparency latent construct and an overlap between confidentiality and control latent constructs in all four scenarios. The items (see~\autoref{tab:questionnaire}) for the transparency construct \textit{t1, t2, t3} show communalities and factor loadings above the recommended thresholds and small cross-loadings. The candidate items for the confidentiality and control construct are \textit{u1, u2, u3, c2}. We excluded \textit{u4, c1, c3 and c4} due to cross loadings and since they do not achieve the commonalities threshold. The details of the EFA results can be viewed in~\autoref{tab:efa} and in Appendix~\ref{app:methods}. Thus, we tested and verified possible combinations of items under these two factors in the MGCFA procedure.

\begin{table}[!ht]
    \centering
    \caption{Exploratory Factor Analysis among the four Scenarios (S) with loadings and communalities ($h^{2}$). Items meeting the thresholds are shown in \textbf{bold}.}
    \renewcommand{\arraystretch}{1.15}
    \begin{tabular}{|>{\centering\arraybackslash}m{1.5cm}|
                >{\centering\arraybackslash}m{3cm}
                >{\centering\arraybackslash}m{1.5cm}|
                >{\centering\arraybackslash}m{3cm}
                >{\centering\arraybackslash}m{1.5cm}|}
        \hline
        \textbf{Item} & \multicolumn{1}{c}{\textbf{S1 MR1 / MR2}} & \textbf{S1 h2} & \multicolumn{1}{c}{\textbf{S2 MR1-MR2}} & \textbf{S2 h2} \\ \hline
        \textbf{u1} & \textbf{0.78 / 0.10} & \textbf{0.72} & \textbf{0.82 / 0.05} & \textbf{0.72} \\ \hline
        \textbf{u2} & \textbf{0.62 / 0.21} & \textbf{0.63} & \textbf{0.63 / 0.20} & \textbf{0.61} \\ \hline
        \textbf{u3} & \textbf{0.90 / -0.16} & \textbf{0.63} & \textbf{0.89 / -0.12} & \textbf{0.67} \\ \hline
        u4 & 0.45 / 0.16 & 0.33 & 0.43 / 0.29 & 0.43 \\ \hline
        c1 & 0.33 / 0.48 & 0.57 & 0.37 / 0.40 & 0.49 \\ \hline
        \textbf{c2} & \textbf{0.71 / 0.19} & \textbf{0.73} & \textbf{0.76 / 0.11} & \textbf{0.70} \\ \hline
        c3 & 0.39 / -0.25 & 0.07 & 0.47 / -0.29 & 0.13 \\ \hline
        c4 & 0.42 / 0.18 & 0.31 & 0.48 / 0.18 & 0.38 \\ \hline
        \textbf{t1} & \textbf{0.02 / 0.86} & \textbf{0.77} & \textbf{0.13 / 0.74} & \textbf{0.69} \\ \hline
        \textbf{t2} & \textbf{0.08 / 0.69} & \textbf{0.57} & \textbf{-0.04 / 0.76} & \textbf{0.55} \\ \hline
        \textbf{t3} & \textbf{-0.01 / 0.84} & \textbf{0.69} & \textbf{0.01 / 0.88} & \textbf{0.78} \\ \hline
        \textbf{Item} & \multicolumn{1}{c}{\textbf{S3 MR1-MR2}} & \textbf{S3 h2} & \multicolumn{1}{c}{\textbf{S4 MR1-MR2}} & \textbf{S4 h2} \\ \hline
        \textbf{u1} & \textbf{0.80 / 0.07} & \textbf{0.71} & \textbf{0.76 / 0.08} & \textbf{0.66} \\ \hline
        \textbf{u2} & \textbf{0.66 / 0.21} & \textbf{0.64} & \textbf{0.51 / 0.30} & \textbf{0.57} \\ \hline
        \textbf{u3} & \textbf{0.83 / -0.04} & \textbf{0.65} & \textbf{0.83 / -0.10} & \textbf{0.58} \\ \hline
        u4 & 0.24 / 0.53 & 0.50 & 0.48 / 0.14 & 0.34 \\ \hline
        c1 & 0.35 / 0.49 & 0.56 & 0.31 / 0.58 & 0.60 \\ \hline
        \textbf{c2} & \textbf{0.90 / 0.00} & \textbf{0.81} & \textbf{0.80 / 0.10} & \textbf{0.76} \\ \hline
        c3 & -0.52 / 0.37 & 0.18 & 0.49 / -0.24 & 0.12 \\ \hline
        c4 & 0.53 / 0.12 & 0.37 & 0.11 / 0.64 & 0.53 \\ \hline
        \textbf{t1} & \textbf{-0.01 / 0.93} & \textbf{0.85} & \textbf{-0.06 / 0.96} & \textbf{0.84} \\ \hline
        \textbf{t2} & \textbf{0.15 / 0.67} & \textbf{0.59} & \textbf{0.10 / 0.71} & \textbf{0.62} \\ \hline
        \textbf{t3} & \textbf{-0.03 / 0.90} & \textbf{0.78} & \textbf{0.01 / 0.83} & \textbf{0.69} \\ \hline
    \end{tabular}
    \label{tab:efa}
\end{table}

Following the steps exemplified in Section~\ref{sec:methods}, we evaluated our questionnaire with the data-driven insights from the EFA by using the MGCFA. Our model was valid and satisfied the recommended metric thresholds by including the items \textit{u1, u2, c2} for Confidentiality and Control's dimension, and items \textit{t1, t2, t3} for the dimension related to Transparency. The four baseline models, each scenario fitted independently one from the other in the CFA model, were obtained with an RMSEA of $0.00$ for Scenario \#1 and \#2 and of $0.03$ for Scenario \#3 and \#4. RMSEA had excellent fits in all the scenarios, with the upper bounds for scenarios \#3 and \#4 that are with mediocre fits and excellent for Scenario \#1 and \#2. The SRMR was in all the scenarios between $0.02-0.03$. Similarly, in all the scenarios (baseline models), the values for CFI and TLI that we obtained were in the range of $0.99-1.00$ for both metrics. All the metrics explored are satisfying and all of them are providing the goodness fit of the models. The internal reliability was measured with Cronbach's alphas ($\alpha$) and found in a range between $0.78-0.84$, considering independently all the scenarios. We integrate the average variance extracted (AVE) that, in all the scenarios, showed the convergent validity to be above the accepted threshold of $0.5$. We measured the HeteroTrait-MonoTrait ratio of correlations (HTMT) to demonstrate discriminant validity which was satisfactory with a value of $0.75$. We further calculated the Congeneric Reliability ($\omega$) that in both confidentiality and control as well as transparency dimensions yielded a score of $0.84$. The configural (scenarios together without constraints), the metric (scenarios together with the fixed factor loadings constraint), the scalar model (scenarios together with fixed factor loadings and item intercepts) and the full model (scenarios together with fixed factor loadings, intercepts and residuals) provided satisfactory results that respected all the metrics' thresholds. Thus, we accepted the full invariance among the four scenarios since the $\Delta CFI \leq 0.01$ between the configural, the metric, the scalar and the full models.  A summary of all the performances related to the metrics for the MGCFA can be seen in ~\autoref{tab:fit_metrics} and are in line with the literature recommendations to assure reliability, validity and global fit of the CFA model~\cite{gross_cfa_2023}. 

These results confirm that our questionnaire consistently measures the intended constructs across all scenarios. Thus, we can continue with our validated measurement tool (e.g., questionnaire) to perform the clustering considering the scenarios that we proposed including the information about their potential privacy risks.

\begin{table}[!ht]
    \centering
    \caption{Multi-Group CFA global fit performances with Configural, Metric, Scalar and Full models regarding the measurement invariance.}
    \renewcommand{\arraystretch}{1.15}
    \begin{tabular}{|
        >{\centering\arraybackslash}m{3cm}|
        >{\centering\arraybackslash}m{4cm}|
        >{\centering\arraybackslash}m{1.25cm}|
        >{\centering\arraybackslash}m{1.25cm}|
        >{\centering\arraybackslash}m{1.25cm}|}
    \hline
     \textbf{Model} & \textbf{RMSEA (CI 90\%)} & \textbf{SRMR} & \textbf{CFI} & \textbf{TLI} \\ \hline
    \textbf{Scenario \#1} & \textbf{0.00 (0.00 - 0.07)} & \textbf{0.02} & \textbf{1.00} & \textbf{1.00} \\
    \textbf{Scenario \#2} &  \textbf{0.00 (0.00 - 0.08}) & \textbf{0.02} & \textbf{1.00} & \textbf{1.00} \\ 
    \textbf{Scenario \#3} & \textbf{0.05} (\textbf{0.00} - 0.10) & \textbf{0.03} & \textbf{0.99} & \textbf{0.99} \\ 
    \textbf{Scenario \#4} & \textbf{0.05} (\textbf{0.00} - 0.09) & \textbf{0.02} & \textbf{0.99} & \textbf{0.99} \\ \hline
    \textbf{Configural} & \textbf{0.03 (0.00 - 0.06)} & \textbf{0.02} & \textbf{0.99} & \textbf{0.99} \\ \hline
    \textbf{Metric} & \textbf{0.03 (0.00 - 0.06)} & \textbf{0.03} & \textbf{0.99} & \textbf{0.99} \\ \hline
    \textbf{Scalar} & \textbf{0.05 (0.03 - 0.07)} & \textbf{0.04} & \textbf{0.99} & \textbf{0.98} \\ \hline
    \textbf{Full} & \textbf{0.05 (0.03 - 0.07)} & \textbf{0.05} & \textbf{0.98} & \textbf{0.99} \\ \hline
    \end{tabular}
    \label{tab:fit_metrics}
\end{table}

\subsection{Privacy Clusters and their Characterization with Profiles}
To answer the research question \textbf{RQ1}, we proceed with the data analysis by employing the hierarchical clustering using the Euclidean metric and the Ward method. The Ward method aims to minimize the increase in variance when merging clusters~\cite{ward_1963_hierarchical}. We obtained the input data for the clustering algorithm from the \textit{lavPredict} function (\textit{lavaan} package in R) which estimates latent response values for each item by utilizing the thresholds corresponding to the ordinal answer options as in~\cite{biselli_questionnaire_2022}. The predicted score values are then derived by calculating the weighted sum of these latent values for each factor, where the factor loadings serve as the weights. 

Since the proven measurement invariance was established across the configural, metric, scalar and full models, we considered the two dimensions (e.g., Control and Confidentiality, and Transparency) across all the scenarios. Thus, we tested 8 parameters (2 dimensions * 4 scenarios) with the number of clusters ($k$) from $2$ to $7$. The optimal $k$ was decided by visualizing the dendrogram and maximising the silhouette score. All together the scenarios reached their maximum values in the silhouette score when the $k=2$ with $0.31$, but considering the dendrograms, we opted for $k=3$, which achieved a silhouette score of $0.21$ To select the number of clusters, we also considered a trade-off between the data-driven approach (induction) as well as the consideration of our goal for a potential real-world implementation (deduction)~\cite{salminen_2022_inddec}. Specifically, the silhouette scores varied by no more than $5\%$ between $k=2$ and $k=3$ considering the $[-1, +1]$ silhouette score range, and the identification of only two clusters would not reflect and capture a broader range of privacy concerns and requirements. More precisely deriving three, instead of two clusters as a basis for expressing data sharing preferences, will allow us to subsequently derive more fine-grained bundles of privacy permission settings for at least three different types of users.

~\autoref{tab:clusters_means_scenarios} shows the results of how participants populated the three clusters in each scenario with related descriptive statistics divided by questionnaire items. We named the three clusters as follows: 
\begin{itemize}
    \item Basic Privacy
    \item Medium Privacy
    \item High Privacy
\end{itemize}

Users are in general demanding transparency and show concerns regarding confidentiality and control - these two dimensions of the questionnaire positively correlate - considering the high score answers from the participants that answered the questionnaire (see~\autoref{tab:clusters_means_scenarios} for items' mean, standard deviation and median aggregated across scenarios), thus privacy concerns and requirements for transparency are basically visible for all privacy clusters and need to be addressed in some form for users in all clusters as important privacy feature (and for this reason, we chose the name "Basic Privacy" instead of "Low Privacy" for the cluster with the lowest privacy concerns and requirements). We did not find any statistically significant evidence in relation to the demographics (including gender, age or region) and the privacy clusters. 

\begin{table}[!ht]
    \caption{Mean, Standard Deviation, and Median per item--- after calculating the mean among the four scenarios--- within each privacy cluster with \%=Number of Participants in percentage.}
    \centering
    \renewcommand{\arraystretch}{1.15}
    \begin{tabular}{|
            p{3.5cm}|
            >{\centering\arraybackslash}m{1cm}|
            >{\centering\arraybackslash}m{1cm}|
            >{\centering\arraybackslash}m{1cm}|
            >{\centering\arraybackslash}m{1cm}|
            >{\centering\arraybackslash}m{1cm}|
            >{\centering\arraybackslash}m{1cm}|}
        \hline
        \textbf{Privacy Level (\% Part.)} & \textbf{u1} & \textbf{u2} & \textbf{c2} & \textbf{t1} & \textbf{t2} & \textbf{t3} \\
        \hline
        \multicolumn{1}{|l|}{\textbf{High Privacy (19\%)}} & & & & & & \\
        Mean   & 4.28 & 4.42 & 4.19 & 4.61 & 4.44 & 4.67 \\
        Standard Deviation    & 0.66 & 0.80 & 0.75 & 0.55 & 0.61 & 0.51 \\
        Median & 4.50 & 4.75 & 4.25 & 4.75 & 4.50 & 5.00 \\
        \hline
        \multicolumn{1}{|l|}{\textbf{Medium Privacy (65\%)}} & & & & & & \\
        Mean   & 3.63 & 3.79 & 3.69 & 4.18 & 4.01 & 4.29 \\
        Standard Deviation    & 0.60 & 0.67 & 0.64 & 0.53 & 0.65 & 0.56 \\
        Median & 3.75 & 4.00 & 3.75 & 4.25 & 4.00 & 4.25 \\
        \hline
        \multicolumn{1}{|l|}{\textbf{Basic Privacy (16\%)}} & & & & & & \\
        Mean   & 3.14 & 3.22 & 3.04 & 3.37 & 3.38 & 3.28 \\
        Standard Deviation    & 0.84 & 0.87 & 0.83 & 0.79 & 0.80 & 0.86 \\
        Median & 3.12 & 3.12 & 3.00 & 3.25 & 3.50 & 3.25 \\
        \hline
    \end{tabular}
    \label{tab:clusters_means_scenarios}
\end{table}

In the \textbf{Basic Privacy cluster} ($16\%$), the participants are overall less concerned than the participants of other clusters and have fewer requirements for transparency, with mean values between $3.04$ and $3.38$ across all dimensions~\autoref{tab:clusters_means_scenarios}). These scores indicate a generally neutral placement, reflecting limited sensitivity to privacy risks. Nevertheless, the cluster still shows a consistent expectation for transparency (\textit{t1–t3}), where the means ($3.28–3.38$) are slightly higher than those for confidentiality and control (\textit{u1–u2-c2}, $3.14–3.22-3.04$). This suggests that even participants with basic privacy concerns value being informed about how their data is handled and have protection and control of their personal data flows in the TAP applications. Among scenarios, the highest scores were observed for Scenario \#2 ("lock the door"), while the lowest was in Scenario \#1 ("sleep log on cloud calendar"), at $3.07$ slightly suggesting that tangible or verifiable actions generate, in comparison, more concerns even in the basic privacy cluster than a nowadays widely adopted solution of connecting health data from smartwatch (e.g., often one of its main functionality) with other smartphone services such as the calendar.

Participants belonging to the \textbf{Medium Privacy cluster} (65\%) have higher concerns and requirements than the ones in the Basic cluster and overall, moderate requirements, with means ranging from $3.63$ to $4.29$. Transparency dimensions consistently scored higher than control and confidentiality. Scenario \#2 ("lock the door") reached the highest scores within this group, while Scenario \#1 ("sleep log on cloud calendar") showed the lowest, suggesting, similar to the Basic Privacy cluster, that physical security actions elicit stronger concerns and requirements than cloud-based automation also in this cluster.

The \textbf{High Privacy cluster} (19\%) is the one with participants that showed overall highest concerns and requirements. The tendency is similar to the other clusters, participants demand more transparency than they are concerned about confidentiality and control. However, their mean scores are less different compared to the other two clusters with all of them exceeding $4.19$ and medians approaching the maximum of the scale. Transparency was the dominant requirement across all scenarios. Regarding the scenarios, all means are above $4.32$. Scenario \#2 again achieved the maximum value within this cluster ($4.63$), while Scenario \#1 recorded the lowest ($4.32$), although still close to the top of the scale. So, participants are concerned about losing control, and they demand transparency about the processing of their data, including how their data will be used and who will use them. The unpacked descriptive statistics per cluster considering the different scenarios can be found in~\autoref{tab:privacy_scenarios_grouped_lines}.

\indent Once we obtained the cluster labels, we trained logistic regression classifiers. In our procedure, the supervised learning algorithms have the goal of predicting the cluster labels from the ordinal scale data directly collected from the participants regarding the attitudinal privacy concerns and requirements asked in the first part of the questionnaire. The Scenario \#1 data can predict the cluster label with $58\%$ of balanced accuracy. Scenario \#2, it is $63\%$, followed by $69\%$ for Scenario \#3 and Scenario \#4. The performances increased when combining two scenarios data to predict the cluster label. The combination of Scenario \#1 and \#2 achieve 80\% of accuracy, Scenario \#3 with \#4 the 75\%, Scenario \#1 and \#3 the 81\%, Scenario \#2 and \#4 the 76\%, Scenario \#1 and \#4 the 73\% and finally Scenario \#2 with Scenario \#3 the 82\%. All scenarios together as input data for the logistic regression classifier reach  85\% of accuracy. Thanks to these logistic regression models, it will be possible to classify a new user into a specific privacy cluster by skipping the transformation employed to get the input data for the clustering algorithms (e.g., MGCFA and \textit{lavPredict} function) and just answer the six questions of our questionnaire in a scenario where information about the privacy risks is provided, while still recommending the use of Principal Component Analysis (PCA) to verify that the developed scales’ items load correctly~\cite{islami2024inter}.

\begin{table}[!ht]
    \caption{Mean, standard deviation, and median for scenarios S1–S4 within each privacy cluster, grouped in scenario pairs.}
    \centering
    \renewcommand{\arraystretch}{1.2}
    \begin{tabular}{|
        >{\centering\arraybackslash}m{2.75cm}|
        >{\centering\arraybackslash}m{1.2cm}|
        >{\centering\arraybackslash}m{1.2cm}|
        >{\centering\arraybackslash}m{1.2cm}|
        >{\centering\arraybackslash}m{1.2cm}|
        >{\centering\arraybackslash}m{1.2cm}|
        >{\centering\arraybackslash}m{1.2cm}|}
        \hline
        \textbf{Privacy Level} & \textbf{Mean} & \textbf{Std} & \textbf{Median} & \textbf{Mean} & \textbf{Std} & \textbf{Median} \\
        \hline
        & \multicolumn{3}{c|}{\textbf{Scenario 1}} & \multicolumn{3}{c|}{\textbf{Scenario 2}} \\
        \hline
        \textbf{High Privacy} & 4.32 & 0.74 & 4.50 & 4.63 & 0.46 & 4.83 \\
        \textbf{Medium Privacy} & 3.61 & 0.71 & 3.67 & 4.17 & 0.61 & 4.17 \\
        \textbf{Basic Privacy} & 3.07 & 0.82 & 3.00 & 3.53 & 0.87 & 3.50 \\
        \hline
        & \multicolumn{3}{c|}{\textbf{Scenario 3}} & \multicolumn{3}{c|}{\textbf{Scenario 4}} \\
        \hline
        \textbf{High Privacy} & 4.40 & 0.73 & 4.58 & 4.38 & 0.61 & 4.50 \\
        \textbf{Medium Privacy} & 3.92 & 0.60 & 4.00 & 4.02 & 0.65 & 4.00 \\
        \textbf{Basic Privacy} & 3.14 & 0.83 & 3.17 & 3.21 & 0.85 & 3.25 \\
        \hline
    \end{tabular}
    \label{tab:privacy_scenarios_grouped_lines}
\end{table}

\indent To answer the \textbf{RQ2}, we further characterized the clusters by including the data obtained in the second part of the questionnaire related to the data sharing preferences. In this regard, we describe the clusters with the frequencies of the isolated features (of the types of data categories, purpose of collection, and data recipient types) from the factorial vignette study procedure according to the related cluster labels. The participants resulted in a distribution where 16\% of them  are in the Basic Privacy Cluster, 65\% in the Medium Privacy Cluster and 19\% in the High Privacy Cluster. 

\begin{figure}[ht]
    \centering
    \includegraphics[width=\linewidth]{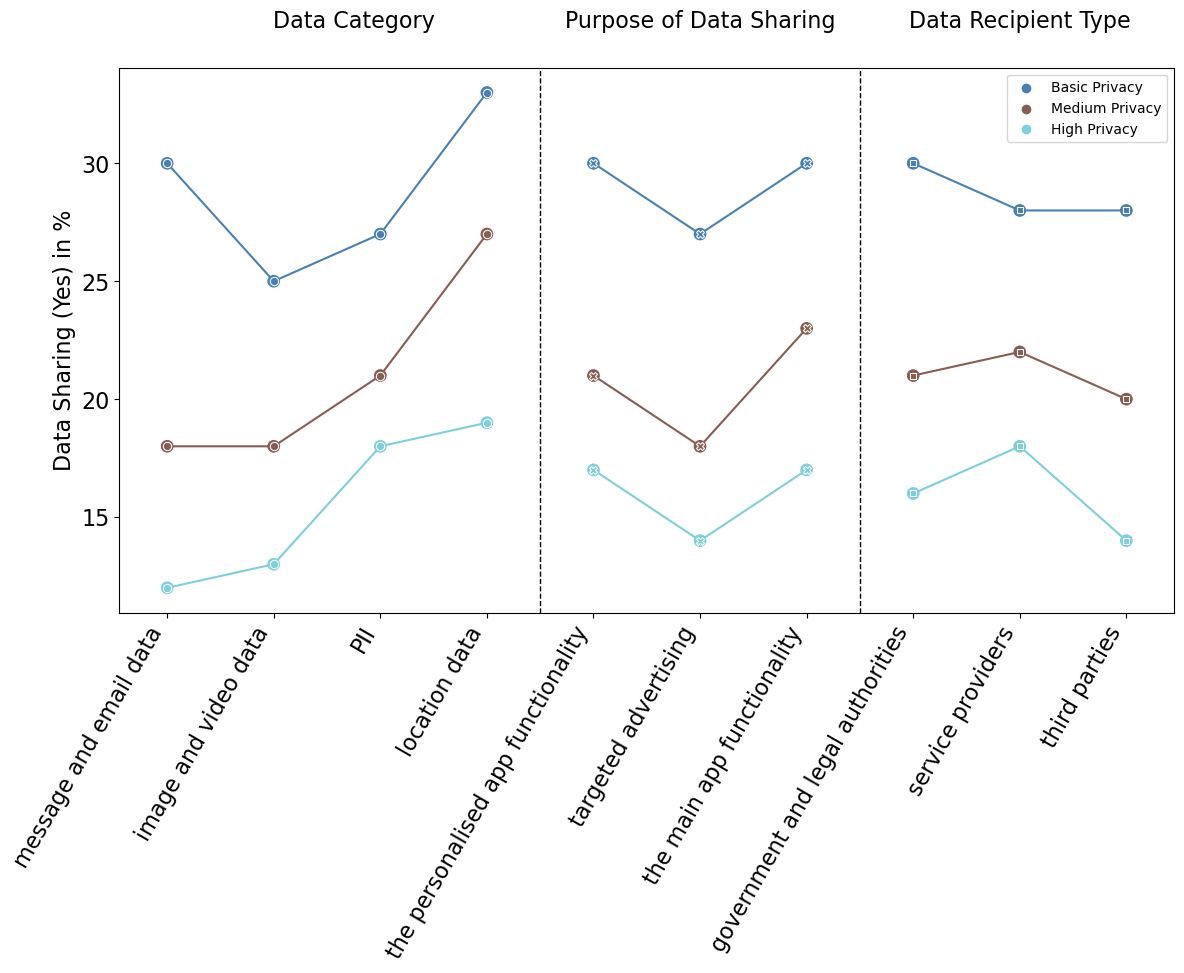}
    \caption{Percentage of Data Sharing in relation to Data Category, Purpose of Data Sharing and Data Recipient Type}
    \label{fig:factorial_trends}
\end{figure}

We plotted the percentage of data sharing to get an intuitive overview of how data category, purpose of data sharing and data recipient type varied among the three clusters (see~\autoref{fig:factorial_trends}). In this second part of the study, we collected higher percentages of no than yes answers related to their potential acceptance to run an IoT app that shares data for all features with yes answers not exceeding 35\% for all features. 
We focused our attention on the differences in the yes answers (i.e. the difference in the acceptance to share data) by characterizing the 3 derived privacy clusters with the privacy data sharing preferences per feature.

As~\autoref{fig:factorial_trends} illustrates, the clustering of users into the three hierarchical privacy clusters (Basic, Medium, High), which we derived from the privacy attitudinal concerns and requirements questionnaire (part 1), is confirmed by the data sharing preferences factorial vignette study procedure (part 2). The factorial vignette study showed that, across the three groups, the preference for potentially accepting data sharing decreased from the High Privacy cluster to the Medium Privacy cluster, and then to the Basic Privacy cluster (even though, as mentioned, users in all clusters rather prefer to not accept data sharing). Correlating the data with each privacy cluster yields corresponding privacy profiles~\cite{inverardi_2023_categorisation}.

Participants belonging to the \textbf{Basic Privacy profile} showed the highest acceptance to share data by running an IoT app, even though the acceptance was still on average $\leq 35\%$ for all features. For this cluster, under the data category feature, location data is the type of data that they prefer to share more than others, followed by message and email data, while image and video data are the data types with the lowest willingness to share. The purpose of data sharing, do not differ among the main app functionality and the personalized one, while the targeting advertising is the less accepted among those. No particular percentage differences were detected regarding the data recipient, with a slightly more willingness to share to government and legal authorities.

The \textbf{Medium Privacy profile} got a lower percentage for a potential acceptance of sharing location data than the Basic privacy cluster and they are less willing to share message and email data with a difference of $13\%$ on average. Putting in a rank the purpose of data sharing sub-levels, higher acceptance of sharing is for the main app functionality followed by the personalized one and ending with targeting advertising. As for the Basic Privacy cluster, no particular differences were found in considering the data recipient type.

The \textbf{High Privacy profile} showed the overall lowest scores compared to the other two clusters, with less than the $\leq 20\%$ acceptance to use an app that shares data for all features on average. The purpose of data sharing had a similar pattern as the Basic Privacy cluster, but with 15\% less of acceptance as expected for the cluster who demand more privacy than the others. The data recipient type had the service providers as higher acceptance to share with followed by the government and legal authorities and finally the third parties.


\section{Discussion}\label{sec:discussion}
The key results in relation to our research questions can be summarized as follows: we show that when participants are informed about the privacy risks of leaked data of our selected IoT TAP scenarios, three privacy clusters (High Privacy, Medium Privacy and Basic Privacy) can be derived for IoT TAPs (\textbf{RQ1}).
To this end, we developed, used and validated a novel questionnaire as a reliable procedure for clustering participants for conducting our study. We further described the clusters with data sharing preferences to extend them in privacy profiles (\textbf{RQ2}) as a step towards finding bundles of privacy settings that could be recommended to users. No evidence was found regarding the impact of the demographics that we collected and the profiles. These key results and their implications will be discussed further in this section. 

Clustering or segmentation processes aim to identify groups of users to better understand their characteristics and needs. A critical step in these methods involves designing a capable measurement tool to collect the necessary data. The data collected by these tools typically includes data collected using questionnaires, behavioral data, or a combination of both~\cite{salminen_persona_2020}. In this work, we validate a questionnaire with two dimensions: Control and Confidentiality, and Transparency to gather privacy concerns and requirements which were analysed to form clusters. These clusters serve as an approximation of the complex decision-making processes involved in setting data sharing preferences. In contrast, current approaches for configuring privacy settings — whether in online services via browsers or on devices through smartphone interfaces — often present users with an abundance of information that they must first comprehend before making informed choices. This can lead to a significant risk of users either making decisions without full awareness due to cognitive overload or lack of knowledge, or even worse by neglecting the process entirely. Thus, the privacy profiles we derived offer an initial framework to guide semi-automated decisions based on users' privacy concerns. Further improving the profiles could later enable individual users to receive recommendations informed by the preferences of similar users. Over time, as users become more familiar with the permission management system, they can additionally personalize their privacy settings or change to a different cluster to adopt more or less stringent privacy controls.

\subsection{Novelty of the Privacy Clusters}\label{sec:disc-noverlty}
Our three privacy clusters were the first privacy clusters that were derived for the context of IoT TAPs for users who were informed about potential risks. Privacy clusters and profiles derived by earlier related work, which we reviewed in Section~\ref{sec:related_work}, were developed for other domains and scopes (e.g. IT or IoT in general and not specifically for IoT TAPs, with users that were not necessarily informed about potential privacy risks and consequences). Also, for this reason, privacy clusters and profiles that were developed earlier for other domains/scopes also differ content-wise from our derived profiles. In particular, the well-known three privacy clusters/categories ("privacy unconcerned", "privacy pragmatists", "privacy fundamentalists") that Alan Westin derived~\cite{westin_1968_privacy}, differ not only domain- and scope-wise but also content-wise significantly from our three privacy clusters (Basic Privacy, Medium Privacy, High Privacy). All participants in our three clusters (including our “Basic Privacy cluster”) share privacy concerns regarding confidentiality and control over IoT TAPs, still to different degrees, possibly also for the reason that our participants were informed about potential risks and consequences. In contrast, users in Westin's cluster “privacy unconcerned” generally do not share privacy concerns and trust organizations processing their personal information and are comfortable with existing organizational procedures. Also, other related works reviewed by~\cite{inverardi_2023_categorisation} that derive three privacy clusters for other domains include, in contrast to our work, a cluster for "privacy unconcerned". Moreover, participants in all three of our clusters are, in general, demanding transparency, a dimension that is not directly captured by Westin’s categories or by most other privacy clusters. 

More details on how our results can facilitate the users' decision-making process regarding IoT TAP permission and controls are discussed in the subsection~\ref{sec:towards_usable} and subsection~\ref{sec:generalize}.

\subsection{Informing on risk}\label{sec:disc-informing}
Our study was based on a questionnaire including IoT TAP application scenarios together with a description of the risks that potential adversaries could infer sensitive information from these applications. 
As discussed above and in line with related work Section~\ref{sec:related_work}, we assume that informing users about privacy risks is relevant for their decision-making~\cite{morgan_delphi_2022} and suggest that IoT TAP users may also need help to understand and reason about privacy risks for developing preferences and concerns~\cite{saeidi_2022_risk}. Moreover, our survey results, which show an overall high requirement for transparency about data processing and storage by participants, are also motivating our approach of informing users about possible personal data inferences that could be made when they decide about permitting the use of their data for an application.

The GDPR also well recognizes the importance of informing users about potential risks in the context of automated decision-making by requiring in its Art. 13 (2) (f) and 15 (1) (h) that data subjects need to be informed about the significance and the envisaged consequences of such processing. While these rules are rather meant for regulating automated decisions made by data controllers without involving the data subjects, our results rely on the literature assumptions (see Section~\ref{sec:intro}) that, even for automation by IoT apps initiated by the users, transparency about privacy risks and their potential consequences is important for users' decision-making when installing apps or setting permissions. 

However, since app providers may not be interested in informing about potential risks and consequences in an unbiased manner, the question of who should provide this risk communication and how it should be provided still needs to be addressed. Another possible direction could be the development of transparency-enhancing tools (TETs) that provide information on potential privacy risks for different types of TAP applications~\cite{breve2022identifying}. TETS can help individuals to exercise their rights for transparency (e.g. pursuant to Art. 12- 15 GDPR), and reduce information asymmetry by providing users with insights about the providers’ data collection and usage and possible consequences that might arise~\cite{murmann,DBLP:journals/corr/Zimmermann15b}. 
A TET providing transparency about risks and potential consequences for IoT TAP applications could become part of or could be invoked by future permission control systems implemented for TAPs. When an IoT TAP application of a certain type is installed or first run by users, or at least at the moment when an action by such an application is triggered that requires to transfer of personal data to a third-party entity, a permission control system at the TAP would request the users for providing their permissions for executing the app or transferring personal data to the third party entity (unless such permissions have already previously set by the users). In this context, the TET could provide or link users to information about potential risks and consequences related to applications of that type. 

\subsection{Towards usable privacy permission and control profiles}\label{sec:towards_usable}
This survey was conducted within the scope of a larger research project on IoT TAP privacy, which has as a future goal to develop usable and transparent privacy control systems. Usability should be achieved via semi-automation with the help of profiles ("bundles") of privacy permission and control settings that can be easily chosen or (semi-)automatically suggested and assigned "on the fly".

Our results show that three hierarchical privacy clusters capturing Basic, Medium and High attitudinal privacy concerns and requirements can be derived for TAP applications for individuals who are informed about potential risks. 
Additionally, the factorial vignette design confirmed that each of the three privacy clusters could be described by one of three distinct attitudinal data sharing preference profiles reflecting users' preferences for Basic, Medium and High privacy protection. 

However, the derived profiles of attitudinal data sharing preferences cannot directly be transferred to different profiles of data sharing permission settings since for all three data sharing preference profiles, users rather preferred not to accept the sharing of their data for different data categories, purposes and data recipient types, even though the extent to which they were accepting data sharing was increasing from profiles that can be associated with the Basic to the Medium and to the High Privacy clusters. 
In compliance with the Data Protection by Default principle of Art. 25 GDPR, users should, by default, be assigned to a Data Protection by Default profile with the most privacy-friendly permission settings. Based on our results, no further profiles can be suggested and offered as alternative profiles which different types of users could then pick.
Nonetheless, our research should be complemented in future with an analysis of behavioral privacy data sharing preferences~\cite{salminen_persona_2020}, especially in the context where users have an interest in using certain types of IoT TAP apps and may therefore be more willing to accept data sharing for being able to use the applications.  This could be done by using the clusters as classification labels to obtain more specific profiles of implementable permission settings that can offer users tailored privacy controls that align with their typical preferences as in~\cite{lipford_2022_privacy}.
As we expect, this may lead to profiles with behavioral data sharing preferences that have on average higher preferences for accepting data sharing than the average attitudinal preferences, and may still capture increasing sharing preferences from Basic to Medium and to High and may thus still be useful for characterizing our three privacy clusters (to be confirmed by our future work). Any profiles with high behavioral preferences/acceptance for data sharing (e.g., above 85\%) for certain items could lead to predefined profiles with permission settings authorizing the data sharing, and these profiles could be chosen and offered to users belonging to the cluster, which the respective preference profile characterizes. Also, machine learning-based privacy assistants, with a design similar to~\cite{das_2018_assistant}, running on the users' (mobile) devices (i.e., under their control) could be implemented and used to evaluate the users’ decisions for developing or adapting privacy profiles for IoT TAPs based on the behavioral user data. Those decisions might have different sources, and it should be up to the user to decide which one to use, for example, from expert opinions or other users' opinions in a collaborative manner~\cite{colnago_2020_ppa, lipford_2022_privacy}
 
Even though our derived attitudinal data sharing preference profiles cannot lead to predefined profiles of data sharing permission settings, our three clusters capturing both increasing privacy concerns, from Basic to Medium and to High, regarding the loss of control as well as increasing transparency requirements, can be used as a basis for deriving and suggesting three profiles of settings related to increasing user control and transparency. We describe these three different profiles based on our findings and adapt them to the actual essence of IoT TAPs which is automation. 

We assume that users in the High Privacy cluster prefer a profile with high privacy controls and low automation. The users in this cluster should therefore always (even if they have already provided informed consent, i.e., set permissions, at the time when the app was installed) be asked to confirm the execution of triggered actions involving certain types of data transfers and processing at third-party servers ("control before final action"), especially if this triggering of actions may impose higher risks (if e.g., it involves personal data that is perceived as sensitive) that they could then be again informed about.

Users in the Medium Privacy cluster find themselves in the middle score profile. Their evaluations have been found with higher variability under the data category and purpose of data sharing, while more stable in terms of data recipient type. They might be asked again to confirm any previously (at installation or run-time) given informed consent in regular intervals when executing apps or sharing their data via IoT TAPs based on the purpose and data category, e.g. with a once-per-week option (or other time intervals suitable for a specific app). Requesting a re-confirmation also allows reminding the users of any potential risks on a regular basis, which they may then be able to contextualise better after having been using the app.

Users in the Basic Privacy cluster, on the other hand, may rather trade high automation and thus more convenience for less control and less privacy. They may thus only be asked for consent for data processing and transfers, and receive information about any potential risks at app installation time. 
 
Privacy profiles deliver solutions that represent groups. Despite that, we interpret these profiles as a starting point for users to eventually adapt, and thus personalise their privacy settings as single individuals~\cite{marky_2024_decide}. Users can receive profile suggestions (with a summary of the profile that suits them best~\cite{marky_2024_decide}), based on the profiles to which they belong, by answering our validated questionnaire in the same setting as our study, which employed the application descriptions and privacy risks.

Privacy profiles can also be semi-automatically developed or updated and personalised "on the fly" depending on how the users behave and interact with consent requests when installing an app, using it for the first time, or when consent needs to be dynamically re-obtained. This will allow us to consider also behavioral privacy preferences. For instance, if a user installs an app "Share my location on Slack", which at certain time intervals posts the user's location on Slack, at app installation, the user consents to share their location. However, if the users’ location becomes sensitive in a certain context (e.g. at specific times when a religious service or political event takes place at a certain location), explicit consent needs to be dynamically requested and obtained according to Art. 9 GDPR. If users, especially those who do not belong to a profile representing the High Privacy cluster, refuse to provide explicit consent, they can be asked if they would like to adapt their permission settings to stricter ones.

\subsection{Too varied to generalize---context-specific scale and profiles as a way forward}\label{sec:generalize}
IoT TAPs have very different characteristics when it comes to different data processing practices, such as the type of data in use, purpose of data usage, data transfer, and data flows between different parties, to name a few and all impact users' data sharing preferences and concerns in this context. Considering the substantial number of IoT TAP applications—numbering in the thousands—it was not feasible to include all possible variations without overwhelming participants or introducing cognitive fatigue. Capturing such vast variability is challenging, especially considering that privacy concerns and preferences are highly contextual. Consequently, in our study, we employed four carefully selected IoT TAP scenarios,  which were formed and filtered based on interview results with experts, see Section~\ref{sec:methods}, to capture a broader spectrum of user concerns and preferences. 
Despite this, our questionnaire is not limited to the scenarios we selected, since during the election period, we considered a widely categorized group of applications in IoT TAPs, and each of the selected scenarios is a representative of a group of scenarios based on certain characteristics, such as personal data sensitivity and data recipient types. For the same reason of avoiding participants' cognitive fatigue, we captured their concerns and preferences using a relatively small number of questions in the questionnaire and features in the factorial vignette study procedure concerning three dimensions of data protection goals that proved to be role-players in users' privacy concerns and preferences.

However, while our questionnaire was designed to be broadly applicable, capturing concerns across key dimensions of data protection goals~\cite{hansen_privacy_2015}, further evaluations are necessary to determine if our developed questionnaire can reliably capture privacy concerns and preferences across various IoT TAP applications, especially those with specific privacy risks not covered in our study~\cite{butori2023construal}. For instance, applications involving technologies like biometric data or those enabling complex data-sharing networks may present unique challenges and privacy risks not addressed in our study. Additionally, these evaluations will help in deriving a comprehensive scale to accurately capture users' preferences and concerns within this context. This will ensure that the resulting profiles remain valid and reliable for a broader range of IoT TAP scenarios. Nonetheless, we should point out that it may not be feasible to generalize privacy concerns and preferences and consequently, the clusters across various ranges of IoT TAPs applications. Further scenario-specific validation is essential for future standardization.

Indeed, data practices within the IoT TAPs context are nuanced and varied, often differing significantly between applications due to factors like technical complexity, regulatory environment, and type of data processed for different purposes. This underscores the importance of tailoring privacy management strategies to specific (groups of) applications rather than assuming a one-size-fits-all approach. It also highlights the need for ongoing monitoring and adaptation of these solutions to accommodate changes in data practices over time~\cite{smith_1996_information}. Ultimately, recognizing the application-specific nature of privacy concerns in IoT TAPs can lead to more effective and user-centric privacy management approaches. Future research should focus on expanding the scope of scenarios and exploring adaptive privacy settings that respond to the evolving landscape of IoT technologies.

\subsection{Limitations}\label{sec:limitations}
First, we recognize that data practices across IoT TAP applications are complex and diverse; therefore, we highlight the need of customizing privacy management strategies to each context rather than applying a one-size-fits-all solution. As discussed in more detail in Section~\ref{sec:generalize}, our results are based on data collected in four IoT TAPs selected scenarios. Although the scenarios represent a range of common characteristics in IoT TAP applications, the complexity and abundance of significantly different data practices and the continual emergence of new technologies and applications mean that additional research could help expand our findings by exploring additional scenarios and adaptive privacy settings that respond to the evolving landscape of IoT technologies.

The results of this study focus on privacy attitudes rather than observable behaviour. While investigating attitudes is relevant—as they reflect how much users care about the processing and sharing of their data—it may not fully represent actual user behaviour. We acknowledge the existence of a privacy paradox, where individuals' behaviours do not align with their expressed attitudes. In the complex IoT context, users often prioritize functionality and convenience over privacy concerns, potentially leading to discrepancies between stated preferences and actions, i.e. privacy paradox.
However, users' concerns and preferences still represent important demands that need to be considered when developing tools, ensuring that privacy settings and protections align with these dispositions. It remains complicated to investigate actual privacy behaviour due to various factors, such as the lack or absence of privacy management tools where to gather the privacy settings or logs like in Android, as well as the vast diversity of applications and contexts in which they are used~\cite{barth_2017_paradox}. These complexities further highlight the importance of understanding privacy attitudes as a foundation for designing systems that can accommodate users’ expectations and concerns.

In the online survey, 301 participants answered the questions. Looking at~\autoref{tab:demographics}, most of our participants were young. Among them, $65\%$ were below $36$ years old. Furthermore, we focused on people living in the EEA and the US, without considering other regions where IFTTT is present. The US and EEA are Western regions with similar basic privacy principles/standards that are enforced by laws (GDPR, US state privacy acts and federal consumer protection laws) and self-regulation and where IFTTT is broadly in use. While this focus ensures some consistency in the legal context, it limits the applicability of our findings to other regions with different cultural attitudes toward privacy and varying regulatory environments. Nonetheless, our survey evaluations showed no evidence that the demographic background, including the region (US, EEA), age or gender, could have any significant impact on our results.

Moreover, 80\% of participants had never used or heard of IoT TAPs like IFTTT. Including non-users was intentional to capture potential future users and avoid bias toward early adopters who might be less privacy-concerned. However, this choice may have influenced the results, as non-users might have different perceptions or heightened concerns due to unfamiliarity with the technology. The hesitation to share personal data observed in our results (see Section~\ref{sec:results}) might be partially attributed to this lack of prior experience. Still, as we discussed above, if we had only conducted the study with IoT TAP users as early adopters, our data set would have likely been biased to represent rather the privacy unconcerned individuals, and hence we made a conscious choice to include a broader range of participants.

Furthermore, the participants were exposed, together with the description of the IoT TAP application, to related potential privacy risks in the first part of the study; thus, this probably impacted their answers. However, as we motivate in Section~\ref{sec:intro} and in Section~\ref{sec:disc-informing}, informing users was an essential choice in our study setup, for enabling users to make informed and conscious decisions.

\section{Conclusion}\label{sec:conclusion}
In an era where IoT TAPs are becoming integral to the personalisation and automation of everyday environments, the challenge of managing privacy effectively has never been more pressing. Our study addresses this issue by exploring how current and potential IoT TAP users' data sharing preferences and concerns can be systematically understood and managed. We showed that when users are informed about privacy risks, their privacy concerns can be clustered using a validated two-dimensional questionnaire focused on confidentiality and control, as well as transparency. This clustering approach was further described using factorial vignette study features into privacy profiles considering data sharing preferences such as data category, purpose of data sharing and data recipient type that linked, as the first step, the privacy disposition to a usable permission and control management system. 

Our findings are based on data collected from four selected IoT TAP scenarios. However, these scenarios include a range of common characteristics within IoT TAP applications, the complexity and number of significantly different data practices, coupled with the continuous emergence of new technologies and applications in the context of IoT TAPs, indicate that additional research could enhance our findings. Therefore, our research suggests that the variability in data sharing preferences and concerns across different IoT TAP applications requires a context-specific approach to privacy management. 
Privacy management permission systems should be tailored to specific (groups of) applications, with ongoing monitoring and adaptation to accommodate changes in data practices and user expectations over time. 

Future work on implementing a comprehensive permission management system is necessary. The next step following this study might involve the development and evaluation of user interfaces or dashboards that allow users, assigned with a profile, to monitor data sharing activities within IoT TAP applications in a transparent way capable to iterate and update dynamically the privacy profiles based on the actual user choices. Such a procedure would implement the privacy recommendations to future IoT TAP users also by showing anonymized comparisons with users of similar profiles. Through the interaction with a transparency dashboard, it may be feasible to incorporate behavioral measures as well to refine, adapt and further validate the privacy profiles based on IoT TAP automation applications review, acceptance or rejection. Additionally, a subsequent phase would focus on enhancing the privacy profiles created in this research by incorporating, through an additional or extended questionnaire, the privacy principles choice and notice/transparency~\cite{iravantchi2025sok} with finer granularity (e.g., how frequently a user wants to be notified, via what channels they want to be notified and where they share their data). These enhancements would assist developers in understanding how frequently each user related to a specific profile prefers to receive notifications and how privacy information should be communicated effectively (see also~\cite{schaub_2015_notice,feng_2021_choice} for the discussion of different choice and notice design spaces). 
Utilizing these clusters as classification labels could enable the creation of more refined privacy profiles including behavioral measures as well as more features or more fine-grained existing features regarding for example sensor types, retention time and data storage location in addition to purpose of collection, the data categories, the recipient type that we used in the current study. 

\subsubsection*{Acknowledgements}
We thank all our participants and those who contributed to the scenario selection. This work was partially supported by the Wallenberg AI, Autonomous Systems and Software Program (WASP) funded by the Knut and Alice Wallenberg Foundation.

\bibliographystyle{unsrt}  
\bibliography{sample-base}
\newpage
\appendix
\section{Introduction and Demographics}
    \begin{tcolorbox}[colback=gray!10, colframe=black!50, title=Useful Information]
    "The Internet of Things (IoT) describes devices with sensors, processing ability, software, and other technologies that connect and exchange data with other devices and systems over the Internet or other communications networks" (Wikipedia definition retrieved Jan 2024). \\ 

    This questionnaire is related to IoT platforms that are online services allowing end-users to connect services and devices through a simple recipe. This recipe follows the formula: IF THIS THEN THAT.
    For example: if you enter a room, then turn on the lights. In this example, the trigger is when the end user enters a room (where a sensor detects presence) and the action is turning on the lights. \\
    
    Between the trigger and action there is an interconnection via the IoT platform. Due to this interconnection, the trigger company provider and the action company provider automatically communicate and know the data related to the app.
    \begin{itemize}
        \item Where are you from? \textit{(Options: European Economic Area, United States of America)}
        \item What's your age group? \textit{(Options: 18-25, 26-35, 36-45, 46-55, 55+)}
        \item What's your gender? \textit{(Options: Woman, Man, Non-binary, Prefer not to say)}
        \item How many IoT devices do you own? \textit{(Options: 0, 1-2, 3-4, 5+)}
        \item How much time per day do you interact with your IoT devices on average? \textit{(Options: 0'-30', 30'-60', 60'-120', 120'+)}
        \item Have you ever had experiences with a data breach (e.g., someone stole your data by hacking an online service)? \textit{(Options: Yes, No)}
        \item Have you ever heard of or used an IoT Trigger-Action platform such as IFTTT, Microsoft Power Automate, or Zapier? \textit{(Options: Yes, No)}
        \item What is your highest level of education? \textit{(Options: Less than High School, High School, University degree)}
    \end{itemize}
    You will be presented with 4 scenarios where such interconnection is used, followed by 11 statements. You will have five response options: strongly disagree, disagree, neutral, agree, strongly agree. After the 4 scenarios, additional yes/no questions will be asked.
\end{tcolorbox}

\section{Study part I: Questionnaire}\label{app:11_questions}
\begin{tcolorbox}[colback=gray!10, colframe=black!50, title=Questionnaire Items]
    Here are the 11 questions of the questionnaire. Note that both the order of questions and the order of scenarios were randomized.
    \begin{enumerate}
        \item I feel that as a result of using this app, my personal information may not remain confidential \textit{(Options: Strongly Disagree, Disagree, Neutral, Agree, Strongly Agree)}.
        \item I am concerned that this app may sell my information to other companies or institutions without my permission \textit{(Options: Strongly Disagree, Disagree, Neutral, Agree, Strongly Agree)}.
        \item I believe that if I use this app, unauthorized people will have access to my data \textit{(Options: Strongly Disagree, Disagree, Neutral, Agree, Strongly Agree)}.
        \item I know that by using this app, an incidental data leakage may happen \textit{(Options: Strongly Disagree, Disagree, Neutral, Agree, Strongly Agree)}.
        \item I would be upset if this app unintentionally triggered and processed my data \textit{(Options: Strongly Disagree, Disagree, Neutral, Agree, Strongly Agree)}.
        \item I am concerned that I may lose control over my data by using this app \textit{(Options: Strongly Disagree, Disagree, Neutral, Agree, Strongly Agree)}.
        \item I feel I have control over my data in this app if the data collection happens in compliance with Privacy Policies, Rules, and Standards \textit{(Options: Strongly Disagree, Disagree, Neutral, Agree, Strongly Agree)}.
        \item For this app, I don't want it to run automatically; I prefer to press a button before the action runs \textit{(Options: Strongly Disagree, Disagree, Neutral, Agree, Strongly Agree)}.
        \item It’s important for me to know what information the service companies involved in this app store about me in their database \textit{(Options: Strongly Disagree, Disagree, Neutral, Agree, Strongly Agree)}.
        \item For this app, I want to receive a summary about the data processing (e.g., how my data are manipulated to produce meaningful information) that occurs \textit{(Options: Strongly Disagree, Disagree, Neutral, Agree, Strongly Agree)}.
        \item It's important for me to know if the data from this app will be sold to third parties for marketing purposes \textit{(Options: Strongly Disagree, Disagree, Neutral, Agree, Strongly Agree)}.
    \end{enumerate}
\end{tcolorbox}

\begin{tcolorbox}[colback=gray!10, colframe=black!50, title=Study Scenarios]
    \textbf{Scenario 1} \\[3pt]
    \textbf{App:} When a new sleep pattern is logged by your smartwatch, this application will add an event in your cloud calendar containing your sleep information. \\[3pt]
    \textbf{Risk:} If an adversary intercepts the communication, there is a risk that sleep routine details may be leaked and used to infer stress levels, fatigue, or even suggest sleeping pills. \\[3pt]
    \textit{Note: Along with the app description, we presented the image~\ref{fig:scenario1} to help participants understand the IoT TAP context.} \\[3pt]
    The 11 questions of the questionnaire are presented here. \\[10pt]
    \textbf{Scenario 2} \\[3pt]
    \textbf{App:} If your location is outside your home, this application will lock the door. \\[3pt]
    \textbf{Risk:} If an adversary collects your location information, they could predict when you are not at home. \\[3pt]
    \textit{Note: Along with the app description, we presented the image~\ref{fig:scenario2} to help participants understand the IoT TAP context.} \\[3pt]
    The 11 questions of the questionnaire are presented here. \\[10pt]
    \textbf{Scenario 3} \\[3pt]
    \textbf{App:} When a new photo is added to your smartphone's photo gallery, this application will upload the photo to your designated cloud storage. \\[3pt]
    \textbf{Risk:} If an adversary gains unauthorized access to the cloud storage, they could extract sensitive information from the uploaded photos. This could include places you visited, timestamps, and people with you, leading to a privacy breach and unintentional exposure of personal details. \\[3pt]
    \textit{Note: Along with the app description, we presented the image~\ref{fig:scenario3} to help participants understand the IoT TAP context.} \\[3pt]
    The 11 questions of the questionnaire are presented here. \\[10pt]
    \textbf{Scenario 4} \\[3pt]
    \textbf{App:} When you like a video on a media platform (e.g., video, photo, news), this application will post the content to your online social network account. \\[3pt]
    \textbf{Risk:} Posting content to an online social network may allow it to target you with advertisements based on your personal preferences from the media platform. \\[3pt]
    \textit{Note: Along with the app description, we presented the image~\ref{fig:scenario4} to help participants understand the IoT TAP context.} \\[3pt]
    The 11 questions of the questionnaire are presented here.
\end{tcolorbox}

\section{Study part II: factorial vignette study}\label{app:factorial_vignette}
\begin{tcolorbox}[colback=gray!10, colframe=black!50, title=Factorial Vignette Study]
    \textbf{Features and Levels:} This part of the study used the features and related sub-levels listed in Table~\ref{tab:practices}. \\[4pt]
    \textbf{Main Question:} \\
    \textit{"Would you accept the following for running an IoT application?"} \\[4pt]
    \textbf{Formula for each of the 36 combinations (4×3×4):} \\
    \textit{Your \{data category\} is shared for \{purpose of data sharing\} and with \{data recipient type\}.} (Options: Yes/No) \\[4pt]
    \textbf{Example:} \\
    \textit{Your location data is shared for the main app functionality and with third parties.} \\[6pt]
    All 36 factorial vignette study applications were presented in random order.
\end{tcolorbox}

\section{Template to Assign a User to a Privacy Clusters}
\begin{tcolorbox}[colback=gray!10,colframe=black!50,title=App Evaluation]
    \textbf{App Description:} If This Then That \\[4pt]
    \textbf{Potential Privacy Risk:} to be manually derived \\[6pt]
    \textbf{Questions (5-point Likert Scale):}
    \begin{itemize}
        \item I feel that as a result of using this app, my personal information may not remain confidential.
        \item I am concerned that this app may sell my information to other companies or institutions without my permission.
        \item I am concerned that I may lose control over my data by using this app.
        \item It’s important for me to know what information the service companies involved in this app store about me in their database.
        \item For this app, I want to receive a summary about the data processing that occurs (e.g., how my data are manipulated to produce meaningful information).
        \item It’s important for me to know if the data from this app will be sold to third parties for marketing purposes.
    \end{itemize}
\end{tcolorbox}

\begin{table}[ht]
    \caption{Study part 1: Original statements from literature that inspired the questionnaire items.} 
    \centering
    \renewcommand{\arraystretch}{1.2} 
    \begin{tcolorbox}[colback=gray!10,colframe=black!50, boxrule=0.6pt, arc=2mm, left=2mm, right=2mm, top=2mm, bottom=2mm, width=\textwidth]
        \begin{tabular}{|>{\centering\arraybackslash}p{2.25cm} 
                        |>{\centering\arraybackslash}p{0.5cm} 
                        |>{\arraybackslash}p{8.5cm}|}
        \hline
        \textbf{Dimension} & \textbf{id} & \textbf{Original Item or Focus Group quotation} \\ \hline
        \multirow[c]{4}{*}{\textbf{Confidentiality}} 
            & u1 & I believe my personal information provided to these Web sites remains confidential~\citep{dinev_privacy_2013}. \\ \hhline{~--}
            & u2 & I am afraid that this data will be used for commercial reasons or that the wrong person will get access to my data. Moreover, I am afraid that insurance companies will use the data to personalize insurances~\citep{maus_privacy_2021}. \\ \hhline{~--}
            & u3 & I believe my personal information is accessible only to those authorized to have access~\citep{dinev_privacy_2013}. \\ \hhline{~--}
            & u4 & Thinking about the possible data leakage, has your desire to keeping using any of these applets changed?~\citep{cobb_ifttt_2020}. \\ \hline
        \multirow[c]{4}{*}{\textbf{Control}} 
            & c1 & Would you be upset if the applet contributed to the following situations occurring: private information gets posted online unintentionally, possibly embarrassing you~\citep{cobb_ifttt_2020}. \\ \hhline{~--}
            & c2 & Thinking about the possible loss of control, has your desire to keeping using any of these applets changed?~\citep{cobb_ifttt_2020}. \\ \hhline{~--}
            & c3 & Importance of whether or not the site posts a privacy policy~\citep{awad_privacy_2006}. \\ \hhline{~--}
            & c4 & I would definitely not like it to upload immediately. It would be very scary if it just uploaded without you like pressing upload~\citep{romare_tapping_2023}. \\ \hline
        \multirow[c]{3}{*}{\textbf{Transparency}} 
            & t1 & Importance of whether a company will allow me to find out what information about me they keep in their databases~\citep{awad_privacy_2006}. \\ \hhline{~--}
            & t2 & Importance of whether a site is going to use the information they collect from me in a way that will identify me~\citep{awad_privacy_2006}. \\ \hhline{~--}
            & t3 & Some Web sites use special identification numbers not only to personalize site content, but also to personalize advertising that appears on the site and make sure that visitors are not repeatedly shown the same advertisements. If a site that you frequented asked you whether it could assign you an identification number so that it could provide you with personalized advertising, would you agree?~\citep{awad_privacy_2006}. \\ \hline
        \end{tabular}
    \end{tcolorbox}
    \label{tab:questionnaire-inspired} 
\end{table}

\section{Methods}\label{app:methods}


\begin{figure}[!ht]
    \centering
    \begin{minipage}[b]{0.45\linewidth}
        \centering
        \includegraphics[width=\linewidth]{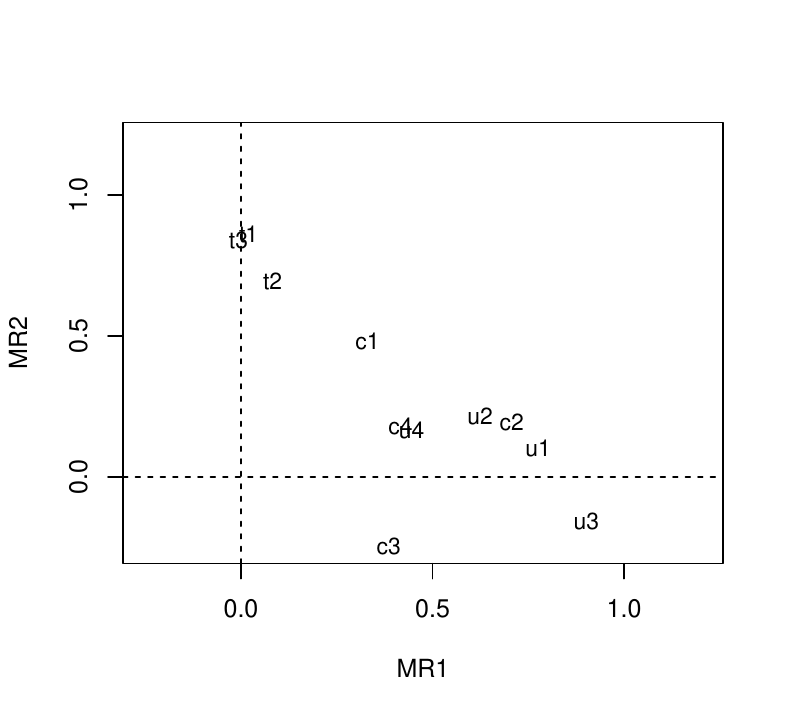}
    \end{minipage}%
    \begin{minipage}[b]{0.45\linewidth}
        \centering
        \includegraphics[width=\linewidth]{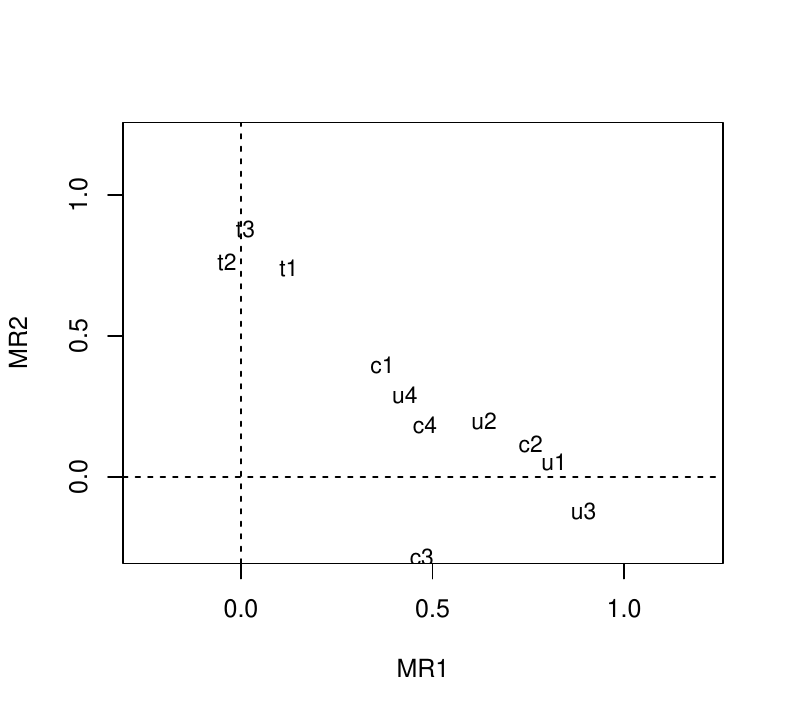}
    \end{minipage}
    \vspace{0.5cm} 
    \begin{minipage}[b]{0.45\linewidth}
        \centering
        \includegraphics[width=\linewidth]{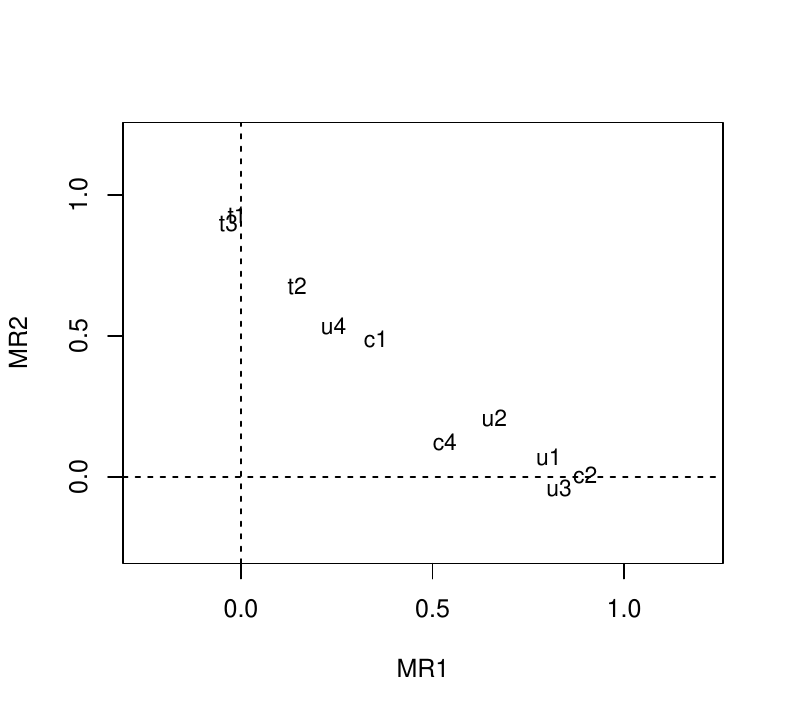}
    \end{minipage}%
    \begin{minipage}[b]{0.45\linewidth}
        \centering
        \includegraphics[width=\linewidth]{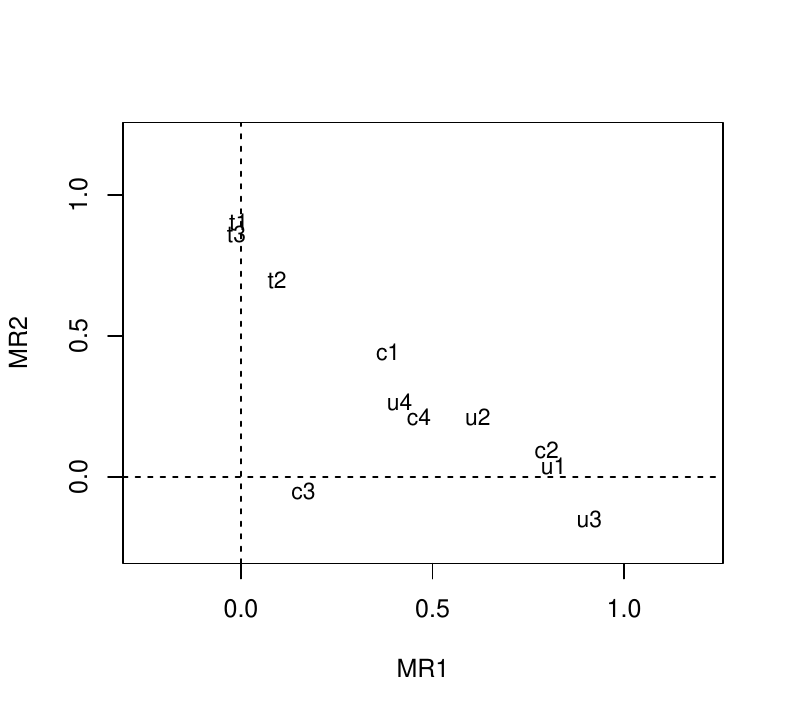}
    \end{minipage}
    \caption{Exploratory Factor Analysis per Scenario, Scenario \#1 (top left), Scenario \#2 (top right), Scenario \#3 (bottom left) and Scenario \#4 (bottom right). The items that can be considered and tested in the MGCFA should load with a value $> 0.32$ and with communalities $h^2 \geq 0.40$, see Table~\ref{tab:efa}.}
    \label{fig:efa}
\end{figure}

\end{document}